\pdfoutput=1
\newif\ifdraft
\newif\iffullversion

\drafttrue

\documentclass[pra, twocolumn,superscriptaddress,  notitlepage, showpacs, floatfix]{revtex4-1}
\usepackage[margin=1in]{geometry}
\usepackage[utf8]{inputenc}
\usepackage[T1]{fontenc}
\usepackage{amsmath,bm,amssymb}
\usepackage{hyperref}
\usepackage{color}
\usepackage{xspace}
\usepackage{graphicx}
\usepackage{enumitem}
\newcommand{\ket}[1]{| #1 \rangle}
\newcommand{\bra}[1]{\langle #1 |}

\newcommand{\bb}[1]{\left( #1 \right)}

\newcommand{\Sz}{\hat{S}_z}

\newcommand{\be}{\begin{equation}}
\newcommand{\ee}{\end{equation}}
\newcommand{\bmul}{\begin{multline}}
\newcommand{\emul}{\end{multline}}
\newcommand{\bea}{\begin{eqnarray}}
\newcommand{\eea}{\end{eqnarray}}

\newcommand{\mat}[2]{
\left[\begin{array}{#1}
#2
\end{array}
\right]}

\renewcommand{\t}[1]{\textrm{#1}}

\ifdraft
\newcommand{\TODO}[1]{{\color{red}[{\bf TODO}: #1]}}
\else
\newcommand{\TODO}[1]{}
\fi

\begin{document}
\title{Many-body effects in quantum metrology}
\author{Jan Czajkowski}
\affiliation{QuSoft, University of Amsterdam, Institute for Logic, Language and Computation (ILLC),\\
P.O. Box 94242, 1090 GE Amsterdam, The Netherlands}
\author{Krzysztof Paw{\l}owski}
\affiliation{Center for Theoretical Physics, Polish Academy of Sciences, Al. Lotników 32/46, 02-668 Warsaw, Poland }
\author{Rafa\l{} Demkowicz-Dobrza\'nski}
\affiliation{Faculty of Physics, University of Warsaw,  ul. Pasteura 5, PL-02-093 Warszawa, Poland}

\begin{abstract}
We study the impact of many-body effects on the fundamental precision limits in quantum metrology. On the one hand such effects may lead to non-linear Hamiltonians, studied in the field of non-linear quantum metrology, while on the other hand  they may result
in decoherence processes that cannot be described using single-body noise models.
We provide a general reasoning that allows to predict the fundamental scaling of precision in such models as a function of the number of atoms present in the system. Moreover, we describe a computationally efficient approach that allows for a simple derivation of quantitative bounds.  We illustrate these general considerations by a detailed analysis of fundamental precision bounds in a paradigmatic atomic interferometry experiment with standard linear Hamiltonian but with \emph{both} single and two-body losses taken into account---a model which is motivated by the most recent Bose-Einstein Condensate (BEC) magnetometry experiments. Using this example we also
 highlight the impact of the atom number super-selection rule on the possibility of protecting interferometric protocols against decoherence.
 \end{abstract}

\maketitle

\section{Introduction}
The research effort in the field of quantum metrology \cite{Giovannetti2011, demkowicz2015quantum, Toth2014, Pezze2016} goes along two main directions. The first is the experimental one,
with the ultimate goal of designing metrological setups that harness quantum behavior of light and matter while at the same time
 minimize unwanted effects of decoherence in order to reach the precision regime inaccessible within the classical paradigm
  \cite{Wasilewski2010, Lucke2011, Ockelon2013, Kruse2016, Schnabel2016}.
The theoretical effort, on the other hand, aims at identifying new promising proposals for quantum metrological protocols as well as  fundamental bounds on precision that can in principle be achieved in quantum systems. The bounds are derived using simplified models that are intended to capture the essential features of the relevant physical systems and take into account the dominant sources of decoherence \cite{Caves1981, Huelga1997, Escher2011, demkowicz2012elusive, knysh2014true,  Sekatski2017, demkowicz2017adaptive, zhou2018achieving}.

Typically, a significant discrepancy is observed between the fundamental bounds arising from theoretical considerations and the performance of practical setups. This is due to either oversimplified theoretical modeling of important physical effects present in the experiment or experimental inability of preparing the optimal quantum states and implementing the optimal quantum operations required by the theory. A noticeable exception from this rule is the case of quantum enhanced optical interferometry, where the performance of  squeezed-light enhanced gravitation wave detector devices \cite{LIGO2011, LIGO2013} approaches closely the fundamental limit predicted by the theory  \cite{Demkowicz2013}.

Atomic systems are inherently more complex than purely optical ones, and hence the theoretical analysis of fundamental metrological bounds
is much more challenging. While in optical interferometry the dominant source of decoherence is loss, which has  single-body character, in atomic systems interactions between atoms naturally lead to various many-body effects. In cold atom experiments, apart from single-particle losses, which in principle might be described using a similar formalism as in the optical case, two- and three-body losses play a crucial role in the system dynamics. This poses a serious challenge to theorists, as the most powerful theoretical methods developed in the field of quantum metrology are naturally tailored to deal with uncorrelated, or in other words, single-body noise \cite{Escher2011, demkowicz2012elusive, Kolodynski2013}.

In this paper, we substantially develop the ideas from \cite{demkowicz2017adaptive}, where it was proposed to view the evolution of an $N$-particle permutationally invariant system with at most $n$-body interactions as equivalent to an application series of $n$-particle quantum channels to $n$-element subsets of $N$ particles. In this way, one is able to make use of theoretical methods, developed with single-body noise models in mind, in many-body noise cases,
without the need to consider the full $N$-particle space but rather a much smaller $n$-particle space. In what follows we will call this the \emph{reduced particle number} (RPN) approach.
In \cite{demkowicz2017adaptive} this idea was used to obtain scalings of precision bounds in cases of a $k$-body Hamiltonian and $l$-body noise processes.
Here we generalize these considerations to a situation where both Hamiltonian and the noise part contains many terms with different level of non-linearity and provide a clear recipe for how to proceed in order to obtain the eventual fundamental precision scaling.
Furthermore, we discuss the way to obtain explicit \emph{quantitative} bounds using this approach, and apply it to find limits on precision of atomic interferometry in presence of both single- and two-body losses.

Interestingly, according to \cite{demkowicz2017adaptive, zhou2018achieving} in case of a linear Hamiltonian the two-body losses are in principle correctable, and therefore one should only focus on the effects of single-body losses. This is true, but we show that this is true only under an  assumption that one is able to prepare states with indefinite particle numbers. Conversely, in case one is limited to definite particle number states, as is the case of atomic interferometry where the particle number super-selection rule holds, this statement is no longer valid. We show this fact analytically, and also derive an explicit bound on precision of linear interferometry in presence of two-body losses. We also show how this super-selection rule should be applied on the level of the RPN approach to obtain explicit quantitative bounds that include the effects of both single and two-body losses that are meaningful for atomic interferometric experiments.
Finally, we compare the obtained bounds with simulations of an explicit atom interferometric protocol as well as data from a Bose-Einstein Condensate (BEC) magnetometry experiment.

The paper is organized as follows. In Section~\ref{sec:methods} we set up the metrological model that serves as a basis for our considerations and review the fundamental tools of quantum estimation and metrology theory that we use in our analysis. In Section~\ref{sec:box-model}, we introduce the RPN approach and provide a general recipe to obtain asymptotic scaling of precision with the number of atoms in models where multiple processes with different degrees of non-linearity act simultaneously. We also show how quantitative bounds can be obtained within the RPN approach.
In Section~\ref{sec:full-losses}, we focus on an atomic interferometry model with linear Hamiltonian in presence of both single- and two- body losses. We show  similarities and differences between the result obtained within the RPN approach and the approach that is based on the study of the properties of operators acting in the full $N$-particle bosonic space, and point to the consequences of application of the atom number super-selection rule.
We compare the bounds with numerical simulations and experimental data. In Section~\ref{sec:conclusions} we conclude the paper and provide an outlook on possible future research directions.

\section{Metrological model\label{sec:methods}}
Let $\rho$ describe the state of an atomic system undergoing an evolution governed by a general quantum master equation \cite{breuer2002theory}:
\begin{align}
\label{eq:master}
\frac{\mathrm{d}\rho}{\mathrm{d}t}= -\text{i}\omega\left[ H,\rho \right] + \underset{\mathcal{L}[\rho]}{\underbrace{  \sum_{j=1}^J L_j\rho L_j^{\dagger}-\frac{1}{2}\rho L_j^{\dagger}L_j- \frac{1}{2}L_j^{\dagger}L_j\rho  }},
\end{align}
where $\omega$, multiplying the generator of the unitary part of the dynamics $H$ (which in what follows we will refer to as the Hamiltonian), is the parameter to be estimated, while $\mathcal{L}(\rho)$ represents decoherence processes, which are assumed here to be Markovian and have no inherent time dependence---formally this means that the dynamics can be described using a semi-group. The goal is to estimate $\omega$ under a fixed total probing time $T$ which is treated here as a resource.
This problem can be viewed as a generalized quantum frequency estimation problem \cite{Bollinger1996, Huelga1997, Macieszczak2014, Smirne2016}.

\begin{figure}
\begin{center}
\includegraphics[width=\columnwidth]{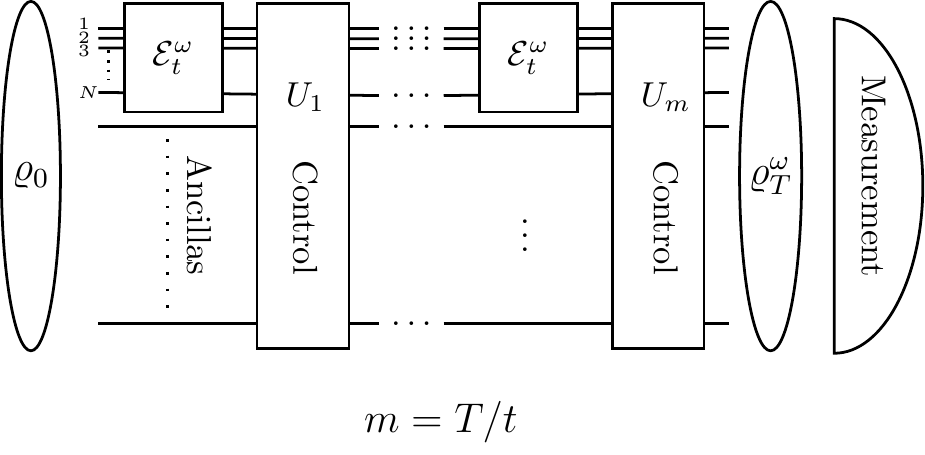}
\caption{The general scheme of an adaptive quantum metrological experiment assuming arbitrary fast control.
$\mathcal{E}_{t}^\omega$ represents physical evolution of the $N$ sensing atoms, while $U_i$ are the control operations that may entangle/disentangle the atoms with arbitrary number of ancillary systems. The goal is to estimate $\omega$ with lowest possible uncertainty under given total time $T$ of the experiment.
\label{fig:ultra-scheme}}
\end{center}
\end{figure}

Let $\mathcal{E}_t^\omega$ be the quantum map representing the above dynamics integrated over time $t$. Hence, if
no other operations are applied to the state we can write the evolved state as
 $\rho_t^\omega = \mathcal{E}_t^\omega(\rho_0)$,
 where $\rho_0$ is the initial state.
Similarly as in \cite{demkowicz2014using, Sekatski2017, demkowicz2017adaptive, zhou2018achieving, laurenza2018channel}  we look for the fundamental bound and therefore consider the most general quantum adaptive metrological protocol, which is depicted in Fig.~\ref{fig:ultra-scheme}.

In the scheme in Fig.~\ref{fig:ultra-scheme} we have explicitly indicated $N$ lines entering each channel, which represent $N$ atoms being used for sensing.
The general scheme allows the atoms to be entangled with an arbitrary number of ancillary systems, which may contribute nontrivially to the sensing process thanks to the unitary control operations $U_i$---we denote the state of the system + ancillas by
$\varrho$ in contrast to $\rho$ which describes the atomic system alone. The scheme, with $N$ separate lines, suggest that we deal with
distinguishable particles. Indeed, our method based on the RPN approach refers to particles as formally distinguishable. Many important atom interferometric experiments, however, are performed with Bose-Einstein condensates, with only the symmetric states involved. Still, since distinguishable particles in principle offer a greater metrological potential, as symmetric states form only a subset of all $N$ particle states, any fundamental bounds derived for distinguishable particles will also be valid for indistinguishable ones. Moreover, these bounds often prove to be saturable by strategies based on the use of symmetric states \cite{demkowicz2015quantum} and hence tightness of the bounds is typically not affected by turning to distinguishable particle paradigm.

Note also, that in the scheme the number of particles $N$ is unchanged from one protocol step to the other. When we deal with losses, this should be understood in the spirit that there are additional internal degrees of freedom of each particle-line that represent the particle being lost. Hence, while the formal number of particles remains unchanged, the loss process can still be described within this framework. This also points to the possibility that control operations $U_i$ may effectively ``reload'' new particles in place of the old lost ones, by a formal change of internal states of each particle-line. We discuss this issue in detail in Sec.~\ref{sec:full-losses}, where we also show how one can tighten the bounds by assuming impossibility of reloading the lost particles on the course of estimation process.

The bound on precision of estimating $\omega$ can be obtained using the Quantum Cram{\'e}r-Rao bound \cite{helstrom1976quantum}:
\begin{equation}
\label{eq:qfi}
\Delta\omega \geq \frac{1}{\sqrt{F_Q}}, \ F_Q = 2 \sum_{a, b} \frac{|\bra{a}\dot{\varrho}^\omega_T\ket{b}|^2}{\lambda_a + \lambda_b},
\end{equation}
where $F_Q$  is the Quantum Fisher Information (QFI), dot signifies $\frac{\t{d}}{\t{d}\omega}$, while $\ket{a}$, $\lambda_a$ are eigenvectors and eigenvalues of the final state of the protocol $\varrho^\omega_T$. Taking QFI as a figure of merit, the goal of determining the fundamental bound can be viewed as the problem of maximizing QFI over input state $\varrho$ and control operations $U_i$. In particular the key qualitative question is whether it is fundamentally possible to protect coherence of the sensing system on long time scales keeping the $T^2$ scaling of QFI---the so called Heisenberg scaling achievable in noiseless quantum metrology---or is the scaling of the QFI bound to be linear in $T$ for sufficiently long times.

This in itself is an extremely complicated mathematical problem, mainly due to the fact that the presence of decoherence causes the state
$\varrho_T^\omega$ to be mixed. Fortunately a powerful theoretical methods based on the idea of minimization over different representations of quantum channels originated in \cite{fujiwara2008fibre} and developed in \cite{Escher2011, demkowicz2012elusive, Kolodynski2013, demkowicz2014using} has finally led to a concise and efficiently computable solution of the problem \cite{demkowicz2017adaptive, zhou2018achieving}, where it was shown that provided the following condition is satisfied
\begin{equation}
\label{eq:space}
H \in \mathcal{S} = \t{span}_{\mathbb{R}}\{\openone, L_j^{\t{H}}, i L_j^{\t{AH}}, (L^\dagger_j L_{j^\prime})^{\t{H}}, i (L^\dagger_{j^\prime} L_{j})^{\t{AH}} \},
\end{equation}
where $^{\t{H}}, ^{\t{AH}}$ denote the Hermitian and anti-Hermitian part of an operator, QFI scales at most linearly with $T$.
We will refer to this condition as the Hamiltonian in the Lindblad Span (HLS) condition.
In this case the explicit upper bound reads \cite{demkowicz2017adaptive, zhou2018achieving}:
\begin{equation}
\label{eq:bound}
\max_{\varrho_0,\{U_i\}} F_Q(\varrho_T^\omega) \leq 4 T \
\min_{\{h,\mathbf{h},\mathfrak{h}\}}  \lVert \alpha \rVert ,  \quad \textrm{s. t.: } \beta=0,
\end{equation}
where $h$ is a real variable, $\mathbf{h}$ a complex vector of length $J$ ($J$ is the number of noise operators $L_j$ appearing in Eq.~\eqref{eq:master}), $\mathfrak{h}$ is a $J \times J$ Hermitian matrix, $\lVert \cdot \rVert$ is the operator norm, while operators $\alpha$ and $\beta$ read
\begin{align}\label{eq:alphabetafull}
\begin{split}
\alpha &= \left(\mathbf{h} \openone + \mathfrak{h}\mathbf{L} \right)^\dagger \left( \mathbf{h} \openone + \mathfrak{h}\mathbf{L}\right ), \\
\beta &  = H  + {h} \openone +  {\mathbf{h}}^{\dagger} \mathbf{L} +\mathbf{L}^\dagger \mathbf{h} +
  \mathbf{L}^\dagger \mathfrak{h} \mathbf{L},
  \end{split}
\end{align}
where the noise operators have been collected in an operator valued vector $\mathbf{L} = [L_1,\dots,L_J]^T$.

If, on the other hand, the HLS condition is not satisfied, one can construct a quantum error-correction protocol that
effectively protects the system from decoherence while preserving some part of the unitary dynamics, which results in an effective
 $T^2$ scaling of QFI \cite{zhou2018achieving}. In most realistic cases, such as cold atom metrology with losses,
 the HLS condition is satisfied and therefore in this paper we will focus mostly on the bound~\eqref{eq:bound}.

If the structure of $H$ and $\mathbf{L}$ is simple enough the above formulation often allows to find an analytical form of the bound \cite{Escher2011, demkowicz2012elusive, Kolodynski2013, demkowicz2017adaptive}.
Otherwise, provided the number of noise operators $\mathbf{L}$ is sufficiently small the above problem can be cast as a simple semi-
definite program and the bound can be found numerically \cite{demkowicz2017adaptive}.

Since we will typically be interested in the large particle number limit, both $H$ and the noise operators $\mathbf{L}$ will
act on a very large Hilbert space and hence in general will not be directly suited to plug into e.g.
the semi-definite program mentioned above. However, in cases where we deal with permutationally invariant systems (such as BEC),
where both $H$ and $\mathbf{L}$ can be described via at most $k$-body interactions, one can effectively calculate the bound
  using only properties of a subchannel that describes the dynamics on a $k$-particle subset of all the particles. Since the dynamics of most cold atom  systems is effectively described with terms that never involve $k$ larger than 3--4, this method will be very efficient numerically and may also help to obtain an analytical form of the bound.  This  approach was introduced in \cite{demkowicz2017adaptive}, and we will refer to it here the \emph{reduced particle number} approach.
   In the next section we will describe this approach systematically and in full generality and
  show how to obtain the fundamental precision scalings in the  general quantum metrology models with many-body effects---the task that was only sketched and studied in some special cases in  \cite{demkowicz2017adaptive}.

\section{Precision bounds via the reduced particle number (RPN) approach\label{sec:box-model}}
 Consider a permutationally invariant system of $N$ particles, where the Hamiltonian $H$ and the noisy part of the dynamics $\mathcal{L}$ can be
 split according to different level of non-linearity of interactions they represent:
 \begin{equation}
H = \sum_k H^{(k)}_N, \quad  \mathcal{L}=  \sum_l \mathcal{L}^{(l)}_N,
\end{equation}
where $H^{(k)}_N$ part contains a contribution to the Hamiltonian arising from $k$-body interactions, and similarly $\mathcal{L}^{(l)}_N$ represents the $l$-body contribution to the noisy part.
More explicitly, permutational invariance of the problem implies that
$H^{(k)}_N = \sum_{\nu \in \Upsilon^k_N} H^{(k)}_{\nu}$, where  $\Upsilon^k_N = \{(i_1,\dots,i_k)\}$ represents all  $k$ element combinations of the $N$ element set, and the operator index $\nu \in \Upsilon^k_N$ denotes the set of particles that a given operator acts on.
Similarly we can write $\mathcal{L}^{(l)}_N = \sum_{\nu \in \Upsilon^l_N} \mathcal{L}^{(l)}_{\nu}$, where $\mathcal{L}^{(l)}_\nu$
 represent the noise part acting on an $l$-particle subset $\nu$.

\begin{figure}
\begin{center}
\includegraphics[width=\columnwidth]{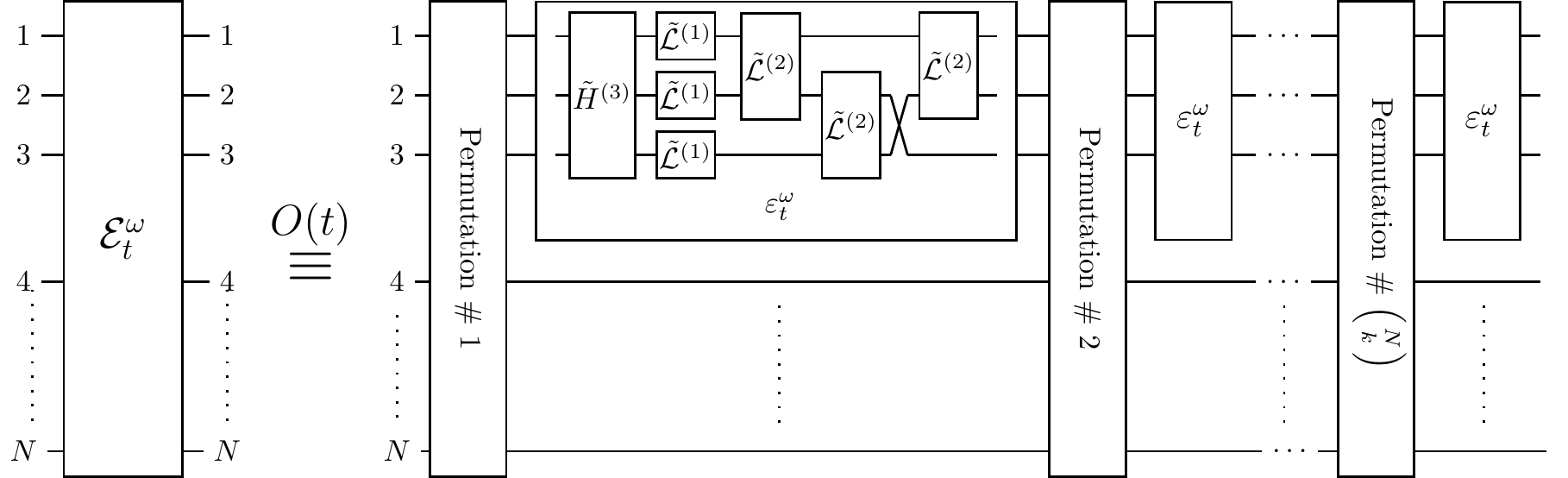}
\caption{The approximation of the quantum channel $\mathcal{E}^{\omega}_t$ in the reduced particle number approach. In this example non-linearity of $H$ is $k=3$ while noisy dynamics $\mathcal{L}$ consists of both single $l=1$ and two-body $l=2$ noise processes.\label{fig:box-scheme}}
\end{center}
\end{figure}

 Since, a duration of a single step in the most general protocol as depicted in Fig.~\ref{fig:ultra-scheme} can without loss of generality be chosen as arbitrarily small, we may equivalently view the action of $H$ and $\mathcal{L}$ as a sequence of operations acting on particular
 sets of particles. According to the Trotter expansion this may at most introduce errors of the order $t^2$.
 Consider for example a situation where we have a three-body $H = H^{(3)}_N$ Hamiltonian and
 both single and two-body noise processes $\mathcal{L} = \mathcal{L}^{(1)}_N + \mathcal{L}^{(2)}_N$.
 We can model the dynamics as effectively consisting of sub-channels $\varepsilon_t^\omega$ acting on all $3$-particle subsets
 of all $N$ particles, as depicted in Fig.~\ref{fig:box-scheme}.

The tilde symbols above the $H$ and $\mathcal{L}$ in Fig.~\ref{fig:box-scheme} are to remind that
due to combinatorial reasons it may be necessary to rescale the operators describing the evolution.
Since we consider a three-particle elementary channel, the three-body Hamiltonian is properly accounted for, there is no need for rescaling and therefore $\tilde{H}^{(3)} = H^{(3)}$, where $H^{(3)}$ represents the $H^{(3)}_{\nu}$ term acting on particles $\nu=(1,2,3)$ as depicted in the picture, and we omitted the $\nu$ subscript for brevity. In what follows whenever we write $H^{(k)}$ or $L^{(l)}$ without any subscript we understand this as a term representing action of Hamiltonian or noise on $k$ or $l$ particle subsets. Now, if we consider $\mathcal{L}^{(1)}$ term, we see that the number of $\tilde{\mathcal{L}}$ gates is
$\binom{3}{1}\binom{N}{3}$, and since there is $N = \binom{N}{1}$ uses of the $\mathcal{L}^{(1)}$ operator in the proper evolution we need to rescale
and set $\tilde{\mathcal{L}}^{(1)} = \binom{N}{1}/\left[\binom{3}{1} \binom{N}{3}\right]\mathcal{L}^{(1)}$.

We can now formulate a general recipe. When considering an RPN model with $n$-particle sub-channels, we need to replace the
original $H^{(k)}$ operators with
\begin{equation}
\label{eq:rescaling}
\tilde{H}^{(k)}_\nu = H^{(k)}_\nu/\chi_k, \quad \chi_k = \frac{\binom{N}{n} \binom{n}{k}}{\binom{N}{k}} \propto N^{n-k}
\end{equation}
and similarly for $\mathcal{L}^{(l)}_\nu$. Note that, since the noisy part $\mathcal{L}^{(l)}_\nu$ is rescaled by a factor of $\chi_l$ the corresponding noise operators $L^{(l)}_{\nu,j}$ that enter quadratically in $\mathcal{L}^{(l)}$ are rescaled as:
\begin{equation}
\tilde{\mathbf{L}}^{(l)}_{\nu} = \mathbf{L}^{(l)}_{\nu}/\sqrt{\chi_l},
\end{equation}
where $\mathbf{L}^{(l)}_\nu$ is a vector containing all noise operators $L^{(l)}_{\nu,j}$ of a given many-body character acting on a given subset of particles $\nu$.

We are now ready to describe a general procedure that allows us to obtain fundamental bounds on precision of estimating $\omega$ in such models.
First, we make the assumption that neither $H^{(k)}$ nor $L^{(l)}$ has any dependence on $N$, so in other words the rates of
elementary processes do not depend on the total number of particles. We will relax this assumption in Sec.~\ref{sec:full-losses}, when discussing the effect of single and two-body losses, where we will take into account the possibility of dependence of two-body loss coefficient on the total particle number via effective changes in the size of the atomic cloud in the trap with the increasing number of atoms.

First of all, if the HLS condition~\eqref{eq:space} is not satisfied, this implies that there is at least one
operator $H^{(k)}_N$ for some $k$ which is not in space $\mathcal{S}$. Hence, one may in principle remove all the noise present in the system via appropriate error-correction protocols and be left with the $N^{2k} T^2$ scaling, known from the models of noise-less non-linear metrology \cite{Boixo2007,luis2007quantum, boixo2008quantum, choi2008bose, napolitano2010nonlinear, gross2010nonlinear, hall2012does, joo2012quantum, sewell2014ultrasensitive}. Motivated by the physical situation in cold atomic systems, in this paper we assume that this is not the case
and all $H^{(k)}_N$ satisfy the HLS condition.

Let us reconsider now the HLS condition in the context of the RPN approach. When considering an $n$-particle subchannel, as in Fig.~\ref{fig:box-scheme}, we can formulate an analogous HLS condition for this subchannel, where the $H^{(k)}$ and $L^{(l)}$ terms
appear in different  variants where they act on different $k$ or $l$ element subsets of $n$ particles.
For example, in the case depicted in Fig.~\ref{fig:box-scheme}, we will ask whether
$H^{(3)}$ that acts on three particles $(1,2,3)$ lives in the space $\mathcal{S}_{n=3}$ defined according to Eq.~\eqref{eq:space} with the following noise operators: single-body operators $L^{(1)}_{\nu,j}$ act on all three particles $\nu=\{1\},\{2\},\{3\}$ and two-body operators
$L^{(2)}_{\nu,j}$ act on all pairs of 3 particles $\nu=\{1,2\}, \{1,3\},\{2,3\}$.
We will refer to the HLS condition on the level of $n$-particle sub-channels as the $\t{HLS}_n$ condition.

Formally, the effective Hamiltonian in the RPN approach can be written as,
\begin{equation}
\tilde{H}_n = \sum_{k} \tilde{H}_n^{(k)}, \quad \tilde{H}_n^{(k)} = \sum_{\nu \in \Upsilon^{k}_{n}} \tilde{H}^{(k)}_\nu,
\end{equation}
while the set of all noise operators combined in a single vector reads:
\begin{equation}
\tilde{\mathbf{L}}_n = \bigoplus_{l} \tilde{\mathbf{L}}_n^{(l)}, \quad \tilde{\mathbf{L}}_n^{(l)} = \bigoplus_{\nu \in \Upsilon_{n}^{l}} \tilde{\mathbf{L}}^{(l)}_{\nu}.
\end{equation}
Hence, the $\t{HLS}_{n}$ condition is given by the same formula as the standard HLS condition given in Eq.~ \eqref{eq:space}, but with $H$ replaced by $\tilde{H}_n$ and the space $\mathcal{S}$ replaced by $\mathcal{S}(\tilde{\mathbf{L}}_n)$, for which the set of noise operators is taken from the vector $\tilde{\mathbf{L}}_n$.

If the $\t{HLS}_n$ condition is satisfied for some $n$ then this implies that the HLS condition is satisfied for the whole system. This simply stems from the fact that full $H$ and noise operators $L_j$ consist of all possible variants of elementary operators acting on different subsets of particles. Therefore if there is linear dependence between operators within a given sub-channel then there will still be a linear dependence if we add all their variants acting on all different subsets. The opposite, however, may not necessarily be true. It might be the case that for some particular $n$ the $\textrm{HLS}_n$ is not satisfied while if we look at the whole system it is. This may be due to the fact that non-trivial new operators appear when we consider products of elementary noise operators $L^{(l)}_{\nu,j}$, $L^{(l^\prime)}_{\nu^\prime, j^\prime}$ with $\nu$ and $\nu'$ only partially overlapping; since such products enter into the definition of the $\mathcal{S}(\tilde{\mathbf{L}}_n)$ space, they may help to span the relevant Hamiltonian.
Still, there is no point in increasing the space of the sub-channel too much, as no qualitatively new operators will appear in the definition of $\mathcal{S}(\tilde{\mathbf{L}}_n)$, if we take $n > 2l$---all possible overlaps of noise operators are then taken into account. Hence when considering the HLS condition using sub-channels one may always limit the considerations to $ \max(k,l) \leq n  \leq \max(k,2l)$---to be precise, here by $k$, $l$ we mean the maximum values of $k$ and $l$ that are relevant in the considered model.

In order to obtain a quantitative bound within the RPN approach using Eq.~\eqref{eq:bound}, note that the number of sub-channels
that appear in the decomposition of $\mathcal{E}^\omega_t$ as depicted in Fig.~\ref{fig:box-scheme} is $\binom{N}{n}$, and therefore
recalling the reasoning that leads to derivation of the bound~\eqref{eq:bound} \cite{demkowicz2017adaptive, zhou2018achieving}---where the number of channel uses is entered as the proportionality factor---the corresponding bound reads:
\begin{equation}
\label{eq:rpnbound}
F_Q \leq 4 T \binom{N}{n} \min_{\tilde{h}, \tilde{\mathbf{h}}, \tilde{\mathfrak{h}}}\| \tilde{\alpha}_n\|, \quad \textrm{s. t.: }  \tilde{\beta}_n=0,
\end{equation}
where
\begin{align}
\label{eq:alphabeta}
\begin{split}
\tilde{\alpha}_n &= \left(\tilde{\mathbf{h}} \openone + \tilde{\mathfrak{h}}\tilde{\mathbf{L}}_n \right)^\dagger \left( \tilde{\mathbf{h}} \openone + \tilde{\mathfrak{h}}\tilde{\mathbf{L}}_n\right ), \\
\tilde{\beta}_n &  = \tilde{H}_n  + \tilde{h} \openone +  {\tilde{\mathbf{h}}}^{\dagger} \tilde{\mathbf{L}}_n +\tilde{\mathbf{L}}_n^\dagger \tilde{\mathbf{h}} +
  \tilde{\mathbf{L}}_n^\dagger \tilde{\mathfrak{h}} \tilde{\mathbf{L}}_n.
  \end{split}
\end{align}
This problem can now be cast as a semi-defnite program as described in \cite{demkowicz2017adaptive} and solved efficiently provided that  $n$ is reasonably small.
First we construct the matrix
\begin{equation}
A_n = \mat{cc}{
\sqrt{\lambda} \openone & \tilde{\mathbf{h}}^{\dagger} \openone + \tilde{\mathbf{L}}_n^\dagger \tilde{\mathfrak{h}}  \\
\tilde{\mathbf{h}} \openone + \tilde{\mathfrak{h}} \tilde{\mathbf{L}}_n & \sqrt{\lambda} \openone^{\otimes J}   },
\end{equation}
where $J$ is the length of the $\tilde{L}_n$ vector.
Minimizing the operator norm $\Vert \tilde{\alpha}_n \Vert$ is now equivalent to minimizing $\lambda$ subject to
$A_n \geq 0$ with the additional constraint coming from the equation $\tilde{\beta}_n=0$.
The bound on QFI can therefore be written as
\begin{align}
\label{eq:semi-defnite}
F_Q \leq  4 T \binom{N}{n}&  \min_{\{\tilde{h},\tilde{\mathbf{h}},\tilde{\mathfrak{h}}\}} \lambda,  \textrm{ subject to: } A_n\geq 0,\  \tilde{\beta}_n=0,
\end{align}
which is indeed a semi-definite program.

Note that the earlier argument stating that increasing $n$ above  $\max(k,2l)$  is not necessary in order to determine
the validity of the HLS condition, has to be carefully reanalyzed when thinking of the actual quantitative form of the bound.
Indeed, increasing $n$ above $\max(k,2l)$ will not change the scaling character of the bound,
 but may sometimes allow us to tighten the coefficient appearing in the bound, see the example of two-body losses in Sec.~\ref{sec:full-losses-rpn}.

\subsection{Scaling in the limit of large particle number}
Without going into numerics, we can get a deeper understanding of
the qualitative behavior of the above bound if we consider the asymptotic limit
of large number of particles $N \rightarrow \infty$.
Recall that according to Eq.~\eqref{eq:rescaling}, the relevant operators scale as $\tilde{H}_\nu^{(k)} \propto N^{k-n}$,
$\tilde{\mathbf{L}}_\nu^{(l)} \propto N^{(l-n)/2}$, hence the terms with larger non-linearity are dominating.

First, for simplicity let us assume that $\tilde{H}_n$ consist of only $k=k^*$-body operators $\tilde{H}_\nu^{(k^*)}$.
Let us also assume that we can satisfy the $\t{HLS}_n$ condition using only noise operators with a given non-linearity $l=l^*$. Then according to Eqs.~\eqref{eq:alphabeta}, in order to satisfy $\tilde{\beta}_n=0$, the corresponding $\tilde{\mathbf{h}}$ coefficients need
 to scale as $N^{k^*-l^*/2-n/2}$ (in order to have $\tilde{\mathbf{h}}^\dagger \tilde{\mathbf{L}}_n$ to scale as $N^{k^*-n}$---the same scaling as the $\tilde{H}_n$), while according to the same reasoning
$\tilde{\mathfrak{h}}$ needs to scale as $N^{k^* - l^*}$ and $\tilde{h}$ needs to scale as $N^{k^*-n}$. Consequently, the resulting $\tilde{\alpha}_n$ will scale as
$N^{2k^{*} - l^{*} - n}$ and finally the bound \eqref{eq:rpnbound} implies
\begin{equation}
\label{eq:scaling}
F_Q \leq \t{const} \cdot  N^{2 k^* - l^{*}}.
\end{equation}
From the above reasoning it follows that if $\tilde{H}_n$ can be spanned by operators with a given type of non-linearity  $l_i$ \emph{on their own},
then we get the tightest bound, in the sense of scaling with $N$, by considering the noise operators with the highest non-linearity. Hence
in the above formula we can write $l^* = \max(\{l_i\})$, where the set $\{l_i\}$ contains non-linearity levels, where for each of them there are corresponding noise operators that allow to satisfy the $\t{HLS}_n$ condition using just operators of a given non-linearity type.

Note that when we have an interplay of decoherence processes with different non-linearities and we investigate the overall scaling of the bound $N^{2 k^* - l^{*}}$ we are left with decoherence processes with relative strength scaling as $\sqrt{N^{l_i-l^*}}$. The tightest bound on QFI will come from $\|\tilde{\alpha}_n\|$ dependent on the strongest decoherence; where by strongest we mean such that scales with $l^*=\max(\{l_i\})$. The reason why we pick this strongest decoherence as the main scaling factor is that only then $\|\tilde{\alpha}_n\|$ depends on asymptotically weakening or constant decoherence processes and those processes that get weaker with increasing $N$ are not crucial to span the Hamiltonian (as all $\mathcal{L}^{(i)}$ are sufficient to span the Hamiltonian on their own).

Consider now, a more complicated situation where in order to satisfy the $\t{HLS}_n$ condition we need to use noise operators with two (or more) different types of non-linearities $\{l_i\}$ \emph{simultaneously}, since none of them on their own is sufficient to set $\tilde{\beta_n} =0$.
In this case, the coefficients $\mathbf{h}$ (and $\mathfrak{h}$) which multiply operators with a given non-linearities $l_i$ will scale
as $N^{k-l_i/2-n/2}$ ($N^{k - l_i}$). When inspecting the asymptotic behavior of $\tilde{\alpha}_n$ we then note that the dominant scaling of $\tilde{\alpha}_n$ with $N$ will be determined by the coefficients $\mathbf{h}$ (and $\mathfrak{h}$) which have the highest scaling exponent---the terms corresponding to noise operators with the lowest non-linearity level. Therefore, in the scaling formula \eqref{eq:scaling}, we need to take
$l^* = \min(\{l_i\})$. The presence of higher order non-linearities, may influence the bound as well, but only in via modification of the multiplying constant and not the scaling---we will see this behavior in \ref{sec:full-losses} when discussing the impact of single and two-body losses on fundamental precision bounds in cold atom metrology.

The argument that bases on the relative strength of decoherence processes does not work in this case. Even though the relative strength grows with $N$ it does not dominate the norm of $\tilde{\alpha}_n$. The part with lower non-linearity is crucial to span the Hamiltonian and so no matter how much the strength of decoherences with higher non-linearities grow, the norm will depend mostly on presence of this lower degree.

So, in short, given Hamiltonian with a fixed non-linearity $k$ we can say that $l^*$ in bound \eqref{eq:scaling} is
equal to the highest level of non-linearity that when combined  with higher non-linearity terms allows to satisfy the $\t{HLS}_n$
condition:
\begin{equation}
\label{eq:lopt}
l^*(k) = \max\left[l: \quad \tilde{H}_n^{(k)} \in \mathcal{S}\left(\bigoplus_{l^\prime \geq l}\tilde{\mathbf{L}}_n^{(l^\prime)}\right) \right].
\end{equation}

Finally, let us allow the Hamiltonian to consist of many terms with different non-linearity levels $k$.
In this case for each $k$-body Hamiltonian part there will be a corresponding $l^*(k)$ providing the tightest bound as given by \eqref{eq:lopt}.
Since all the Hamiltonian parts are present simultaneously, the resulting scaling will be determined by the part that dominates the
formula for $\tilde{\alpha}_n$. This will be the term with non-linearity $k$ for which $2k - l^*(k)$ is maximal.
Hence, in general, the scaling of precision resulting from the RPN approach is given by the formula \eqref{eq:scaling} with
\begin{equation}
k^* = \t{argmax}_k[2k - l^*(k)], \ l^*=l^*(k^*).
\end{equation}
This bound resolves the open problem of scaling of metrological bounds analyzed in \cite{delcampo2017nonlinear, braun2018quantum}.

\subsection{Adapting the bound to experimental realities}
While the preceding considerations take into account experimental imperfections (general Markovian noise),
there are some implicit assumptions that were made on the way to the formula for the fundamental bound \eqref{eq:bound} that
might be modified in order to take into account additional experimental limitations present in the physical systems considered.

First of all, throughout the derivation we have assumed a fixed number of particles $N$ being present in the system.
It does not mean that we exclude losses (they can be formally incorporated via additional internal degrees of freedom of the particles), but it means that in principle, the adaptive strategies are allowed to reload the particles continuously during the course of the experiment.
In real cold atom experiments, this is not the case, and only after the experiment with a given cloud of atoms is finished a new sample is prepared. If we want our bound to represent this experimental limitation we may do it by taking into account  the
time dependence of the number of atoms in the experiment $N(t)$. Given this function, we may easily adapt the reasoning leading to the bound~\eqref{eq:bound} following the general recipe discussed for time dependent models in \cite{demkowicz2017adaptive}, and obtain the corresponding bound
\begin{equation}
\label{eq:rpntbound}
F_Q \leq 4 \int_{0}^T \t{d}t\,  \binom{N(t)}{n} \min_{\tilde{h}, \tilde{\mathbf{h}}, \tilde{\mathfrak{h}}}\| \tilde{\alpha}_n(t)\|, \quad \tilde{\beta}_n(t)=0,
\end{equation}
where the dependence of $\tilde{\alpha}_n$, $\tilde{\beta}_n$ on $t$ is due to the rescaling procedure that leads from non-tilded quantities to tilded quantities involving the particle number parameter $N$, which now is time dependent.

The other potential modification in the preceding derivation may be due to the fact that the noise parameters have some additional nontrivial dependence on the particle number on their own. As discusses in detail in Sec.~\ref{sec:full-losses} this is a typical situation in cold atom systems, where with increasing number of particles the size of the atomic cloud increases and effectively the two-body loss coefficient decreases.  In such situations, this effect needs to be taken into account in order to provide the proper scaling exponent. Still provided that noise coefficient
go down slower than $N^{-1}$ the hierarchy of non-linearities remains valid, in the sense that higher order non-linear terms dominate over lower order non-linear terms in the limit of large $N$---see Sec.~\ref{sec:full-losses} for more detailed discussion in the context of cold atom interferometry experiments.

\section{Quantum interferometry with single- and two-body losses\label{sec:full-losses}}
In this section we present the full analysis of the simultaneous effects of single- and two-body losses on quantum interferometric protocols, keeping the cold atom physical context in mind as the main motivation for this study.
Consider an atomic interferometry model with a system of $N$ two-level atoms where the dynamics is described by the following master equation:
\begin{equation}
\frac{\t{d} \rho}{\t{d} t} =  - \text{i} \omega \left[H^{(1)}, \rho\right] + \mathcal{L}^{(1)}(\rho) + \mathcal{L}^{(2)}(\rho),
\end{equation}
where
\begin{equation}
H^{(1)} = \frac{1}{2}(a_1^\dagger a_1 -a_2^\dagger a_2)
\end{equation}
is the standard linear interferometry Hamiltonian, with $a_i$ ($i=1,2$) being the annihilation operators removing an atom from level $i$,
while the noise operators corresponding to the single and two-body losses parts $\mathcal{L}^{(1)}$ and $\mathcal{L}^{(2)}$ read respectively
\begin{equation}
 L^{(1)}_i = \sqrt{\gamma_i} a_i,\quad  L_{ij}^{(2)}= \sqrt{\gamma_{ij}}a_i a_j.
\end{equation}
In \cite{demkowicz2017adaptive} it was stated that since two-body loss operators and their products are linearly independent from $H^{(1)}$, the system may be effectively protected from two-body losses via some kind of quantum error-correction protocol, and the resulting bound on precision will be determined solely by single-body losses in the case of the linear Hamiltonian.
This reasoning made an implicit assumption, however, that we put no additional restriction on the allowed class of states. We show below that if we  impose a super-selection rule that forbids preparation of superpositions of states with different total atom numbers, and if at least two out of the three two-body loss coefficients $\gamma_{ij}$ are nonzero, it is no longer possible to protect the system against two-body losses. As a result, in the large particle limit the actual bound will be determined by the two-body effects which leads to a much less favorable scaling of precision than the one resulting from the impact of single-body losses only. 

\subsection{Impossibility to protect against two-body losses under atom number superselection rule}
Following the general considerations presented in \cite{demkowicz2017adaptive, zhou2018achieving}, in order to protect the sensing system from two-body losses we need to construct at least a two-dimensional code space, $\{\ket{\psi_1}, \ket{\psi_2}\}$, where the following conditions need to be satisfied:
\begin{align}
\bra{\psi_k} L_{\mathbf{j}}^{(2)\dagger} L_{\mathbf{j}^\prime}^{(2)}  \ket{\psi_{k^\prime}} = \delta_{kk^\prime} \mu_{\mathbf{j}\mathbf{j}^\prime},
\end{align}
where the bold index $\mathbf{j} \in \{ (1,1), (1,2), (2,2), 0 \} $ represents the two-body loss operator double indices, where value of the index $0$ returns the identity matrix $L_0^{(2)} = \openone$, and $\mu_{\mathbf{j}\mathbf{j}^\prime}$ is some Hermitian matrix.
Furthermore, in order to have a nontrivial effective action of the Hamiltonian in the code space, and hence be able to sense the parameter of interest, we require that
\begin{equation}
\bra{\psi_k} H^{(1)} \ket{\psi_{k^\prime}} \neq \t{const} \cdot \delta_{kk^\prime}.
\end{equation}
Consider now
\begin{multline}
\delta_{kk^\prime}  \mu_{ii,ii}= \bra{\psi_k} L^{(2)\dagger}_{ii} L^{(2)}_{ii} \ket{\psi_{k^\prime}} = \gamma_{ii} \bra{\psi_k} a_i^{\dagger 2} a_i^2 \ket{\psi_{k^\prime}} = \\
\gamma_{ii} \bra{\psi_k} a_i^{\dagger} a_i a_i^\dagger a_i   -  a_i^\dagger a_i \ket{\psi_{k^\prime}}.
\end{multline}
We now assume that the code states have a definite atom number, and hence
$(a_1^\dagger a_1 + a_2^\dagger a_2) \ket{\psi_k} = N \ket{\psi_k}$---note that the atoms may still be entangled with some additional ancillas, and hence $\ket{\psi_k}$ may in principle live on a larger Hilbert space, but the assumption simply means that within the system Hilbert space we will only make use of states with a fixed number of particles in both modes. Using this assumption we get
\begin{align}
\label{eq:qec1}
\delta_{kk^\prime} \mu_{11,11} &= \gamma_{11} \bra{\psi_k}a_1^\dagger a_1 (N-1) - a_1^\dagger a_1 a_2^\dagger a_2 \ket{\psi_{k^\prime}} \\
\label{eq:qec2}
\delta_{kk^\prime}  \mu_{22,22}&= \gamma_{22} \bra{\psi_k}a_2^\dagger a_2 (N-1) - a_1^\dagger a_1 a_2^\dagger a_2 \ket{\psi_{k^\prime}}.
\end{align}
Provided $\gamma_{12}>0$ the error-correction condition for the $L^{(2)}_{12}$ operator yields
\begin{equation}
\label{eq:qec3}
 \bra{\psi_k}a_1^\dagger a_1 a_2^\dagger a_2 \ket{\psi_{k^\prime}} = \frac{1}{\gamma_{12}}\delta_{kk^\prime} \mu_{12,12}.
\end{equation}
As a result we have:
\begin{equation}
\bra{\psi_k} a_{1}^\dagger a_1 \ket{\psi_{k^\prime}} = \frac{\delta_{kk^\prime}}{(N-1)} \left(\frac{\mu_{11,11}}{\gamma_{11}} + \frac{\mu_{12,12}}{\gamma_{12}}\right)
\end{equation}
and analogously for $\bra{\psi_k} a_{2}^\dagger a_2 \ket{\psi_{k^\prime}}$. This implies that
\begin{equation}
\bra{\psi_k} H^{(1)} \ket{\psi_{k^\prime}} = \frac{\delta_{kk^{\prime}}}{N-1}\left(\frac{\mu_{11,11}}{\gamma_{11}} - \frac{\mu_{22,22}}{\gamma_{22}}  \right)
\end{equation}
and hence the effective Hamiltonian is trivial on the codespace. The above derivation is valid provided all $\gamma_{ij} >0$.
However, since $(a_1^\dagger a_1 + a_2^\dagger a_2) \ket{\psi_k} = N \ket{\psi_k}$ even if one of the coefficients is zero
two out of three error correction (\ref{eq:qec1},~\ref{eq:qec2},~\ref{eq:qec3}) conditions are sufficient to arrive at the same conclusion.

If on the other hand two coefficients are zero the reasoning fails and this is the only situation when one can construct an error correcting code to protect the sensing process from two-body losses with the atom number superselection rule imposed.

To give a concrete example. Assume $\gamma_{11}>0$ while $\gamma_{12}=0$, $\gamma_{22}=0$.
In this case a possible codespace (which may be shown to be optimal, using the criterion given in Eq.~(36) in \cite{zhou2018achieving})
consists of vectors $\ket{\psi_1} = s \ket{N}\ket{0} + \sqrt{1-s^2} \ket{0} \ket{N}$,
$\ket{\psi_2} = \ket{N/2} \ket{N/2}$, where $s=\sqrt{\frac{N-2}{4(N-1)}}$, and we assumed $N>4$.
The only nontrivial error correcting condition to check is
$\bra{\psi_1} a_1^{\dagger 2}a_1^2 \ket{\psi_1} = \bra{\psi_2} a_1^{\dagger 2}a_1^2 \ket{\psi_2}$ which indeed holds  thanks to the choice
of the parameter $s$ as above. Furthermore, the Hamiltonian is nontrivial in the codespace, since
$\bra{\psi_1} H^{(1)} \ket{\psi_1} = -\frac{N^2}{4(N-1)}$, while all other matrix elements of $H^{(1)}$ in this subspace are zero.
In the end this leads to QFI $F_Q = \frac{T^2 N^4}{(N-1)^2 16}$, which yields the scaling as in the ideal lossless case but with
 a quantitative reduction by approximately a factor of $16$. Similar constructions can be performed when a different loss coefficient is nonzero.

As mentioned above, if two or more two-body loss coefficients are non-zero it is not possible to construct the correcting protocol with the superselection rule imposed. If one insists on constructing such a code one must violate the particle number superselection rule.
For example, a code that works in case all types of two-body loss processes are present, can be written as a simple generalization of the code presented above: $\ket{\psi_1}= (s \ket{N} + \sqrt{1-s^2} \ket{0}) \ket{N/2}$, $\ket{\psi_2}=\ket{N/2}  (s \ket{N} + \sqrt{1-s^2} \ket{0})$.
It can be easily checked that, provided $N>4$, it satisfies error-correcting conditions with respect to all types of two-body loss processes and yields a non-trivial Hamiltonian within the codespace.

\subsection{Asymptotic bounds due to two-body losses}\label{sec:asymptotic}
Assuming that at least two $\gamma_{ij}$ coefficients are non-zero, we will now use the formula~\eqref{eq:bound} to
derive the bound on performance of linear interferometry due to the presence of two-body losses. Assuming that we are in the large $N$ limit, and $\gamma_{ij}$ coefficient are constant with growing $N$ (or declining slower than $N^{-1}$) this will also be a valid bound in presence of \emph{both} single- and two-body losses, since as discussed in detail in Sec.~\ref{sec:methods} the two-body effects will dominate over single-body ones in the limit of large $N$.

The procedure will be similar to that presented in \cite{demkowicz2017adaptive} where the bounds due to two-body losses have been derived, but with a
 notable exception that here we focus on the linear Hamiltonian rather than the quadratic one.

The $\beta$ quantity from Eqs.~\eqref{eq:alphabetafull} reads:
\begin{multline}
\beta = \frac{1}{2}(a_1^\dagger a_1 - a_2^\dagger a_2) + h \openone + \mathfrak{h}_{11,11} \gamma_{11} a_1^{\dagger 2}a_1^2 +  \\\mathfrak{h}_{22,22} \gamma_{22}  a_2^{\dagger 2}a_2^2
+\mathfrak{h}_{12,12} \gamma_{12} a_1^\dagger a_2^\dagger a_1 a_2 + \dots,
\end{multline}
 where we neglected the terms for which the coefficients are set to zero as they do not help in satisfying the condition $\beta=0$.
 We keep in mind, that the operator will be acting within the space of states with total fixed atom number $N$, hence
 we can use the relation $a_1^\dagger a_1 + a_2^\dagger a_2 = N \openone$ provided these operators will act directly on the potential states (in other words are not acted upon by other operators in the formula).
 As a result we can write:
 \begin{align}
 \beta &= \openone\left(h - \frac{N}{2} + N(N-1) \gamma_{22} \mathfrak{h}_{22,22}\right) + \nonumber \\
 & a_1^\dagger a_1 \left[1 +(N-1)(\gamma_{11} \mathfrak{h}_{11,11} - \gamma_{22}\mathfrak{h}_{22,22})\right] +\nonumber \\
 & a_1^\dagger a_1 a_2^\dagger a_2 \left(\mathfrak{h}_{12,12} \gamma_{12} -\gamma_{11}\mathfrak{h}_{11,11}  -\gamma_{22}\mathfrak{h}_{22,22} \right) =0.\label{eq:betah1l2}
 \end{align}
 Consequently, we can express all coefficients as a function  of a single parameter $h$:
 $\mathfrak{h}_{11,11}= - \left(1/2 + \xi  \right)/(N \gamma_{11})$,
 $\mathfrak{h}_{22,22}=  \left(1/2 - \xi  \right)/(N \gamma_{22})$,
 $\mathfrak{h}_{12,12}= - 2 \xi/(N \gamma_{12})$, where  $\xi = \frac{h}{N}$, and we approximated $N-1 \approx N$ as we are focused on the large $N$ regime.
 In order to obtain the bound we now need to minimize the operator norm of
 \begin{align}
 \alpha = &\frac{(1/2+\xi)^2}{N^2\gamma_{11}}a_1^{\dagger 2}a_1^2 +  \frac{(1/2-\xi)^2}{N^2 \gamma_{22}}a_2^{\dagger 2}a_2^2 +\nonumber \\
 &\frac{4\xi^2}{N^2 \gamma_{12}} a_1^\dagger a_2^\dagger a_1 a_2
 \end{align}
over the parameter $\xi$. All involved operators are diagonal in the atom number basis $\ket{n}\ket{N-n}$. Hence the operator norm of $\alpha$ equals the largest absolute value of its diagonal elements in this basis and as a result the formula for the bound reads (again approximating $N-1\approx N$ when necessary):
\begin{multline}
F_Q \leq 4 T \min_\xi \max_{0 \leq x \leq 1} \\ \frac{(\frac{1}{2} +\xi)^2}{\gamma_{11}} x^2 + \frac{(\frac{1}{2} -\xi)^2}{\gamma_{22}}(1-x)^2 + \frac{4 \xi^2}{\gamma_{12}}x(1-x),
\end{multline}
where we will treat $x=n/N$ as a continuous parameter in what follows. This optimization can be easily done numerically, as this function involves at most products of quadratic terms in $x$ and $\xi$.
We can also extract an analytical formula from this expression with some additional assumptions on the relation between $\gamma_{ij}$
coefficients.

With fixed $\xi$ the bound is a quadratic function of $x$. Hence the maximum over $x$ is achieved on
either the boundaries of the region $x=0$, $x=1$ or at the extremum point.
If for a given $\xi$ the function $x$ is convex in $x$ then the maximum is indeed achieved on the boundaries.
This is the case provided $4 \xi^2/\gamma_{12} \geq (\xi +1/2)^2/\gamma_{11} + (1/2 -\xi)^2/\gamma_{22}$.
Assuming this condition is satisfied we can write:
\begin{multline}
F_Q \leq 4T \min_\xi \max\left\{\frac{(1/2+\xi)^2}{\gamma_{11}}, \frac{(1/2-\xi)^2}{\gamma_{22}}\right\}
\end{multline}
where the minimum is achieved for $\xi = \frac{1}{2}\frac{\sqrt{\gamma_{11} - \sqrt{\gamma_{22}}}}{\sqrt{\gamma_{22}}+\sqrt{\gamma_{11}}}$.
The above reasoning is valid provided the function is indeed convex in $x$ for the optimal $\xi$ parameter found above.
Plugging it into the condition for convexity we arrive at the condition for validity of the above bound
$\gamma_{12} \geq  \frac{1}{2} \left(\sqrt{\gamma_{11}}-\sqrt{\gamma_{22}}\right)^2$, and the final form the bound reads:
\begin{equation}
F_Q \leq  \frac{4 T}{(\sqrt{\gamma_{11}} + \sqrt{\gamma_{22}})^2}, \quad \gamma_{12} \geq  \frac{1}{2} \left(\sqrt{\gamma_{11}}-\sqrt{\gamma_{22}}\right)^2.
\label{eq:fullaabound}
\end{equation}
 In particular, the above abound is always valid in the case of symmetric losses $\gamma_{11}=\gamma_{22}$. Note that the bound is similar to the one derived for single-body losses \cite{Kolodynski2013,demkowicz2017adaptive},
but differs by the missing factor $N$, which is in agreement with our general scaling considerations which imply that for $k=1$ Hamiltonian and $l=2$ noise-type QFI should approach a constant value for large $N$.

The other case that is easy to do analytically and is also relevant in experimental context (see Sec. \ref{sec:full-losses-rpn}) is the case where one of the two-body loss coefficients is zero. Let us take $\gamma_{22}=0$ and $\gamma_{12},\gamma_{11}>0$. In this case the minimization over $\xi$ becomes trivial because $\gamma_{22}=0$ implies $\xi=-1/2$, note the discussion below Eq.~\eqref{eq:betah1l2}. We are left with the task of calculating $\max_{0 \leq x \leq 1} \frac{(1-x)^2}{\gamma_{11}} + \frac{x(1-x)}{\gamma_{12}}$. This quadratic function decreases monotonously on $0\leq x\leq 1$ for $\gamma_{11}\leq 2\gamma_{12}$, hence the maximum is attained for $x=0$. In the case when $\gamma_{11}>2\gamma_{12}$ the function is concave with the extremum point in $x=\frac{2 \gamma_{12}-\gamma_{11}}{2 (\gamma_{12}-\gamma_{11})}$. Finally the bound reads
\begin{equation}
F_Q \leq \begin{cases} \frac{4T}{\gamma_{11}}, & \gamma_{11}\leq 2 \gamma_{12} \\
\frac{T\gamma_{11}}{\gamma_{12}(\gamma_{11}-\gamma_{12})}, & \gamma_{11} > 2 \gamma_{12}
\end{cases}
  \label{eq:fqloss2full}.
\end{equation}

\subsection{The reduced particle number (RPN) approach}
\label{sec:full-losses-rpn}
We will now show how the problem of deriving the bound---taking into account single and two-body losses---can be analyzed within the RPN approach,
and discuss the advantages compared with the reasoning presented in the previous subsection.
If we would like to derive a bound in the regime of $N$ where contribution of single and two-body is comparable we would not be able to
follow the simplified reasoning presented in the previous subsection. Instead we would need to take into account all single and two-body loss operators, consider the condition for $\beta=0$ and minimize the operator norm of $\alpha$. This would be cumbersome, first of all due to a larger number of independent parameters over which we need to optimize and furthermore  when calculating the operator norm we need to face the fact that the operators act in a very large Hilbert space.  The RPN approach will allow us to reduce the problem to a small dimensional Hilbert space, which allows us to effectively perform numerical optimization required for the calculation of the bound even if the procedure involves optimization over many free parameters.

We follow here the RPN approach using two-particle $n=2$ sub-channels. Apart  from the two internal states $\ket{1}$, $\ket{2}$ representing the particle being in mode $1$ or $2$ respectively, we also introduce a vacuum state to represent the particle being lost $\ket{v}$, and hence
the two-particle space is $9$ dimensional.
Within the space of two particles we will also restrict ourselves
to symmetric states, as this is the class of states that we encounter in cold atom systems, and moreover, we have numerically checked that relaxing this requirement did not have any impact on the derived bounds.
The symmetric subspace is $6$ dimensional and is spanned by $\mathcal{H}^S_2:=\{ \ket{0,0},\ket{0, 1 },\ket{1, 0 },\ket{0 ,2 }, \ket{1, 1},\ket{2,0} \}$,
where we use the standard occupation number notation with numbers representing the number of particles occupying mode $1$ and $2$ respectively.
     The relation between these basis states and the states  of distinguishable particles with three internal degrees is straightforward. For example: $\ket{0,0} = \ket{v}\otimes \ket{v}$, $\ket{1,0 }=\frac{1}{\sqrt{2}}(\ket{1}\otimes\ket{v}+\ket{v} \otimes \ket{1})$,  $\ket{0,2} = \ket{2} \otimes \ket{2}$,  etc.

We can now write the appropriately rescaled Hamiltonian and noise operators
\begin{align}
\label{eq:operatorsRPN}
\begin{split}
\tilde{H}_2^{(1)} &= \frac{1}{2(N-1)}\left( \tilde{a}_1^\dagger \tilde{a}_1 - \tilde{a}_2^\dagger \tilde{a}_2\right), \\
\tilde{L}^{(1)}_{2,i} &= \sqrt{\frac{\gamma_i}{N-1}} \tilde{a}_i, \quad \tilde{L}^{(2)}_{2,ij} = \sqrt{\gamma_{ij}} \tilde{a}_i \tilde{a}_j,
\end{split}
\end{align}
where the annihilation operators restricted to the considered subspace are given by:
\begin{align}
\tilde{a}_1:= \ket{0,0}\bra{1,0}+\ket{0,1}\bra{1,1}+\sqrt{2}\ket{1,0}\bra{2,0},\\
\tilde{a}_2:= \ket{0,0}\bra{0,1}+\ket{1,0}\bra{1,1}+\sqrt{2}\ket{0, 1}\bra{0,2}.
\end{align}
Since all operators now act in a $6$-dimensional space one can calculate the bounds efficiently using the semi-definite program defined in~\eqref{eq:semi-defnite}.

Without resorting to numerics, using simple algebra of $6\times 6$ matrices one can also analyze the implications of $\tilde{\beta}_2=0$
condition. First, of all one notices that while $\tilde{H}_2 \in \mathcal{S}(\mathbf{\tilde{L}}^{(1)})$ is inside the space spanned by
single-body loss operators, it is not spanned by the the two-body loss operators $\tilde{H}_2 \notin \mathcal{S}(\mathbf{\tilde{L}}^{(2)})$.
This is analogous to the situation we have encountered when analyzing the full operators acting on the whole Hilbert space without the atom-number superselection rule imposed. Indeed, this is to be expected, since in the way we approached the problem using the RPN formalism we in fact allow for states which are superposition of different atom number states up to the total atom number $n=2$. If we want to derive tighter bounds using the RPN approach, that take into account the atom number superselection rule we must formulate the problem mathematically in a way that forbids the use of superposition of different atom number states.

The simplest approach, is to modify the operators
in Eqs.~\eqref{eq:operatorsRPN} in a way that their input space is restricted to the subspace with exactly two-atoms,
so $\mathcal{H}^S_{2!} = \{ \ket{0 ,2 }, \ket{1, 1},\ket{2,0} \}$. As discussed earlier, physically this would correspond to the situation where, at every adaptive step of the protocol lost atoms are being replaced with new ones, so that the total number of atoms is unchanged and we can consider only states with definite number of atoms throughout the protocol.
Formally, this assumption can be incorporated in the RPN approach by restricting the input space of
noise operators $\tilde{L}^{(1)}_i$, $\tilde{L}^{(2)}_{ij}$ to $\mathcal{H}^S_{2!}$:
${\tilde{L}}^{(1)}_{2!} = \tilde{L}^{(1)}_2 P_{\mathcal{H}^S_{2!}}$, where $P_{\mathcal{H}^S_{2!}}$ is the projection on $\mathcal{H}^S_{2!}$,
as well as projecting the Hamiltonian on this subspace ${\tilde{H}}_{2!}^{(1)} = P_{\mathcal{H}^S_{2!}}\tilde{H}_2^{(1)}P_{\mathcal{H}^S_{2!}}$. Note, that we are multiplying noise operators only from one side, as this is how they act on the state of the system, while we project the Hamiltonian from both sides. By doing so, we get that
${\tilde{H}}_{2!}^{(1)}$ is effectively a $3 \times 3$ matrix acting on $\mathcal{H}^S_{2!}$ and so are the products of noise operators and their Hermitian conjugation ${\tilde{L}}_{2!,j}^\dagger {\tilde{L}}_{2!,i}$ (the single noise operators that also appear in the definition of the $\mathcal{S}(\tilde{\mathbf{L}}_{2!})$ space will not be relevant as they introduce terms that are beyond the space spanned by the Hamiltonian and hence need to be set to zero in order to satisfy $\tilde{\beta}_{2!} = 0$).

The operators restricted to the $\mathcal{H}^S_{2!}$ subspace which are relevant for derivation of the bound read explicitly:
\begin{align}
\tilde{H}^{(1)}_{2!} &= \frac{1}{N-1}\mat{ccc}{-1 & 0 & 0 \\ 0 & 0 & 0 \\ 0 & 0 & 1}, \\
\frac{\tilde{L}^{(1)}_{2!,1}}{\sqrt{\frac{\gamma_1}{N-1}} } &= \mat{ccc}{0 & 0 & 0 \\ 0 & 0 & 0 \\ 0 &0 & 0\\0 & 1 & 0\\0 &0 & \sqrt{2}\\0 & 0 & 0}, \frac{\tilde{L}^{(1)}_{2!,2}}{ \sqrt{\frac{\gamma_2}{N-1}}} = \mat{ccc}{0 & 0 & 0 \\ 0 & 0 & 0 \\ 0 &0 & 0\\\sqrt{2}& 0 & 0\\0 & 1 & 0\\0 & 0 & 0},\nonumber \\
 \frac{\tilde{L}^{(2)}_{2!,11}}{\sqrt{\gamma_{11}}} &=   \mat{ccc}{0 & 0 & 0 \\ 0 & 0 & 0 \\ 0 &0 & 0\\0& 0 & 0\\0 & 0 & \\0 & 0 & \sqrt{2}}, \frac{\tilde{L}^{(2)}_{2!,12}}{\sqrt{\gamma_{12}}} =   \mat{ccc}{0 & 0 & 0 \\ 0 & 0 & 0 \\ 0 &0 & 0\\0& 0 & 0\\0 & 0 & \\0 & 1 & 0},\nonumber \\
\frac{\tilde{L}^{(2)}_{2!,12}}{\sqrt{\gamma_{22}}} &=   \mat{ccc}{0 & 0 & 0 \\ 0 & 0 & 0 \\ 0 &0 & 0\\0& 0 & 0\\0 & 0 & \\ \sqrt{2} & 0 &0}.
\end{align}
It is now clear, that provided two of the two-body loss coefficients are non-zero $\tilde{H}^{(1)}_{2!} \in \mathcal{S}(\mathbf{\tilde{L}}^{(2)}_{2!})$ and hence we recover the observation that we have arrived at while considering the operators on the whole $N$-particle Hilbert space: with the super-selection rule imposed the two-body losses will determine the asymptotic scaling in the limit of large $N$. Here we see it since the two-body noise terms are not renormalized by $1/(N-1)$ factor and hence will be dominating in the large $N$ limit.

\begin{figure}
\begin{center}
\includegraphics[width=\columnwidth]{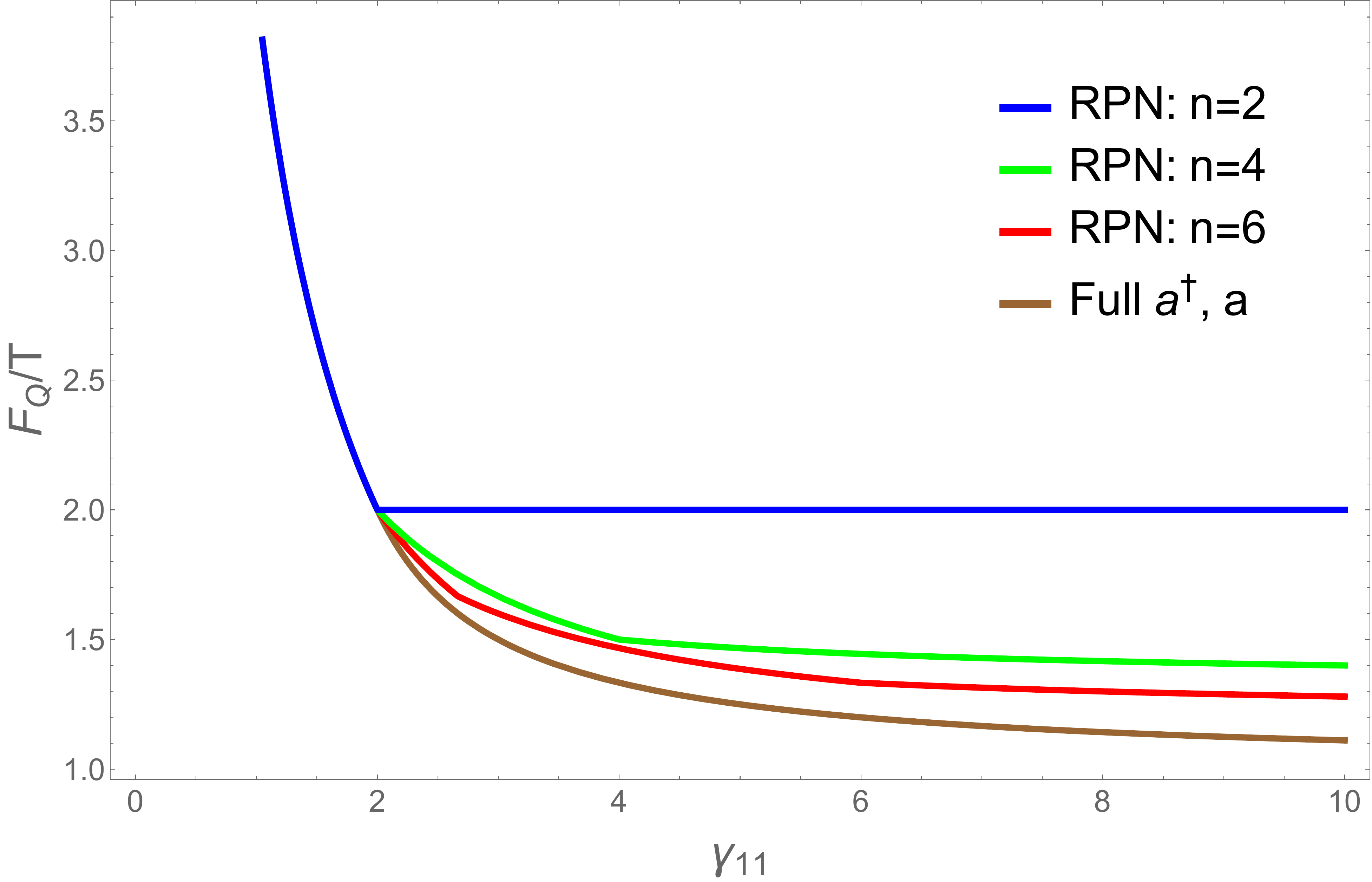}
\caption{Comparison of bounds on QFI in the case of the single-body Hamiltonian and two-body losses with full anihilation operators and in the RPN model. Here we set $\gamma_1=\gamma_2=\gamma_{22}=0$. Bounds are qualitatively the same, two-body losses are enough to set $\beta=0$ but differ quantitatively: the more particles we include in the RPN model the better bound we get.  \label{fig:comparison}}
\end{center}
\end{figure}

In the case considered in Eq.~\eqref{eq:fullaabound} (so in articular symmetric losses) the bound achieved in the RPN model is the same no matter what $n$ we take. However, if we consider an experimentally motivated case $\gamma_{22}=0$, $\gamma_{12}>0$, and $\gamma_{11}>0$, we see that we can tighten the bound in some regime of parameters by increasing $n$, see  Fig.~\ref{fig:comparison}. The bigger space gives us freedom to consider states with the different ratio of atom occupation  numbers in modes 1 and 2, which is useful to tighten the bound in some parameter regions.
So in this case, it appears that while, qualitative scaling of the bound is seen in the RPN approach already at the smallest $n=2$ subchannel considerations, the tightness of the bound may sometimes be improved by increasing $n$.

\subsection{Comparison of the bounds with simulations and experimental results}
The physical system particularly relevant to theoretical considerations is the Bose-Einstein condensate.
Here we briefly recall basics of the bimodal Bose-Einstein condensates in the context of metrology. It will give us better understanding of the experimental situation necessary to make a meaningful comparison of the presented bounds with experimental results.

We focus on the case with $\ket{1}$ and $\ket{2}$ corresponding to internal atomic states \cite{riedel2010, gross2010nonlinear}, although there are also experiments on different spatial modes of a double well potential \cite{esteve2008, Maussang2010, Berrada2016} and proposals to use different orbitals of a scalar fragmented Bose-Einstein condensate \cite{fraisse2018}.
It is instructive to consider first two clouds of atoms, one with $N_1$ atoms in  an internal state $\ket{1}$ and the second with $N_2$ atoms in another internal state $\ket{2}$. The Bose-Einstein condensation means that the total state of the system is close to a pure separable state, built up from two macroscopically occupied orbitals, $\psi_{\rm GPE}^{(1)}$ and $\psi_{\rm GPE}^{(2)}$:
\begin{align}\label{eq:product-state}
\psi (\bm{r}_1,\ldots,\bm{r}_{N_1} ;\, \bm{r}_{N_1+1},\ldots,\bm{r}_{N_1+N_2} ) = \nonumber \\
 \prod_{i=1}^{N_1} \, \psi_{\rm GPE}^{(1)} (\bm{r}_i)\, \prod_{i=N_1}^{N_1+N_2} \, \psi_{\rm GPE}^{(2)} (\bm{r}_i).
\end{align}
The formula relies on the fact, that the atoms in the internal states $\ket{1}$ are distinguishable from the atoms in the state $\ket{2}$. The symmetrization has to be performed between atoms in $\ket{1}$, and between atoms in $\ket{2}$, but not necessarily between atoms in different internal states. In practice, when the evolution starts with a single Bose-Einstein condensate with all atoms initially in $\ket{1}$, shined with the Rabi pulse to obtain the spin coherent state, the state is totally symmetric. These subtle differences, do not impact the further considerations.
It follows from the theory of condensation, that the two orbitals are minimal energy solutions of coupled Gross-Pitaevskii equations  \cite{ho1996, Esry1997}
\begin{align}\label{eq:GPE-coupled}
\mu_1& \psi_{\rm GPE}^{(1)}  = - \frac{\hbar^2 \, \Delta}{2m} \psi_{\rm GPE}^{(1)} + U_1 \psi_{\rm GPE}^{(1)}+ \nonumber \\
		& \bb{g_{11} N_1\left| \psi_{\rm GPE}^{(1)}\right|^2 + g_{12} N_2\left| \psi_{\rm GPE}^{(2)}\right|^2 } \,\psi_{\rm GPE}^{(1)} \,,\\
		\mu_1 & \psi_{\rm GPE}^{(2)}  = - \frac{\hbar^2 \, \Delta}{2m} \psi_{\rm GPE}^{(2)} + U_1 \psi_{\rm GPE}^{(2)}+ \nonumber \\
		& \bb{g_{22} N_2\left| \psi_{\rm GPE}^{(1)}\right|^2 + g_{12} N_1\left| \psi_{\rm GPE}^{(1)}\right|^2 }\,\psi_{\rm GPE}^{(2)} \,,
\end{align}
where coupling constants $g_{\epsilon \epsilon'}$ refer to the interaction between two atoms, one of which is in the state $\ket{\epsilon}$, and the other in $\ket{\epsilon'}$ and , $m$ is the atomic mass of the atoms. The eigenvalues $\mu_{\epsilon}$  are Lagrange multipliers, which physically are the chemical potentials of condensates. Finally, functions $ U_{\epsilon}(\bm{r})$ are the state-dependent potentials trapping atoms
	\begin{equation}
		U_{\epsilon}(\bm{r}) = \frac{1}{2}m\sum_{\sigma=x,\,y,\,z} \omega_{\sigma\epsilon}^{2} \bb{r_{\sigma} - r_{\sigma}^{(\epsilon)}}^2,
	\end{equation}
where $ \omega_{\sigma\epsilon}$ are the trap frequencies, and $r_{\sigma}^{(\epsilon)}$ are trap centers for the atoms in the state $\ket{\epsilon}$.
	
The energy of the two condensates is, up to a constant, given by the formula
	\begin{align}\label{eq:E-GPE}
		& E_{\rm GPE} = \sum_{\epsilon=1,2} N_{\epsilon}\int {\rm d}^3r  \bb{\psi_{\rm GPE}^{(\epsilon)}}^{*}\cdot \nonumber\\
		& \bb{-\frac{\hbar^2 \Delta}{2m} + U_{\epsilon} + \frac{1}{2} \sum_{\epsilon'} g_{\epsilon  \epsilon' }N_{\epsilon'} \,|\psi_{\rm GPE}^{(\epsilon')}|^2} {\psi_{\rm GPE}^{(\epsilon)}}.
	\end{align}
In a typical experiment the initial state is a spin coherent state which is a superposition with the total number of atoms differently distributed between two condensates. The evolution of such system
 (in the states for which dispersions of occupations	are small, precisely $\Delta N_1 \ll N$ and $\Delta N_2 \ll N$)
 is approximately generated by the operator
	\begin{equation}
		\hat{H} = \chi \Sz^2 + \tilde{\chi}\,\hat{N}\,\Sz + u(\hat{N}),
		\label{eq:ham-oat-simple}
	\end{equation}
where $\Sz := \bb{\hat{N}_1 - \hat{N}_2}/2$ is the imbalance between condensates' occupations.
Parameters of the Hamiltonian \eqref{eq:ham-oat-simple} have to be derived from the solutions of the Gross-Pitaevskii equations and the corresponding energies according to \cite{sinatra1998}:
	\begin{align}\label{eq:chis}
		\chi =&\left.  \frac{1}{2\hbar}\,\bb{\frac{\partial^2 E_{\rm GPE}}{\partial N_1^2}+\frac{\partial^2 E_{\rm GPE}}{\partial N_2^2}-2\frac{\partial^2 E_{\rm GPE}}{\partial N_1\,\partial N_2}}\right\rvert_{N_i=\bar{N}_i},\\
		\tilde{\chi} =&\left. \frac{1}{2\hbar}\,\bb{\frac{\partial^2 E_{\rm GPE}}{\partial N_2^2} - \frac{\partial^2 E_{\rm GPE}}{\partial N_1^2}}\right\rvert_{N_i=\bar{N}_i}.
	\end{align}
The derivatives with respect to occupations $N_1$ and $N_2$ are evaluated around the average occupations, i.e. $\bar{N}_1$ and $\bar{N}_2$, which in the initial state are equal to $N/2$.
The central term of the Hamiltonian \eqref{eq:ham-oat-simple} is the one-axis twisting model, $\chi \Sz^2$. Hence, one expects that the system can be used to obtain entangled states, in particular states useful in metrology. Indeed, the spin-squeezed state of two Bose-Einstein condensates have been produced  already a decade ago \cite{esteve2008, riedel2010, gross2010nonlinear} and used to demonstrate a gain in precision of an interferometer.

On the other hand, the potential benefits from condensation may be limited due to particle losses, which are an important source of decoherence for Bose-Einstein condensates. Particle losses are usually divided into three classes, respectively of the number of atoms which are lost in a single event. One-body losses are caused by  collisions between condensed atoms and other particles, from the residual gas unpumped from the vacuum chamber in which the condensate is kept.
			
\begin{figure}
	\includegraphics[width=\columnwidth]{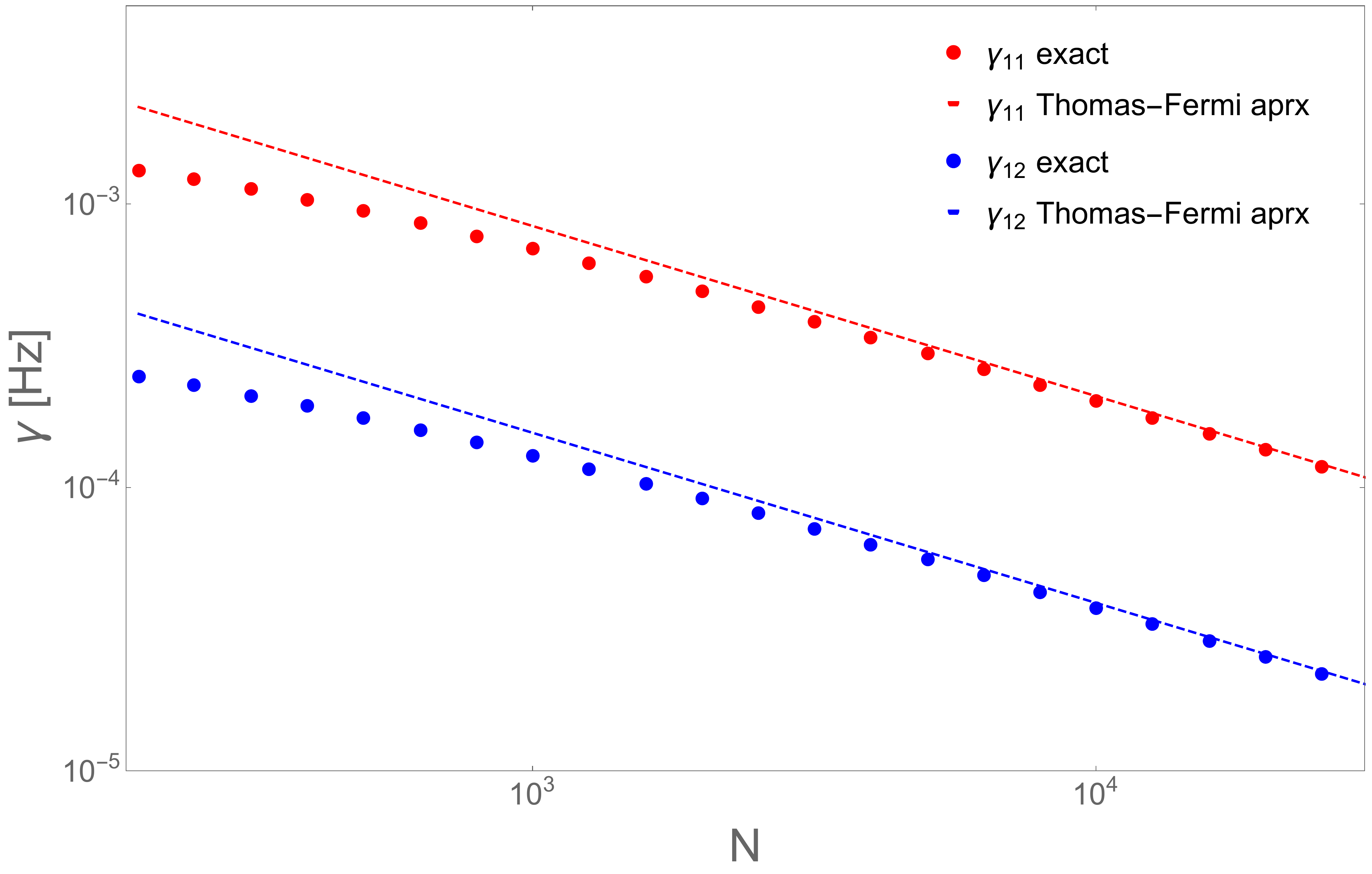}			
	\caption{Rates $\gamma_{11}$ and $\gamma_{12}$ of two-body losses as a function of the total number of atoms $N$. The dashed lines are obtained within Thomas-Fermi approximation (Eq. \eqref{eq:gamma-TF}) and compared with the exact values, evaluated from the numerical solution of the coupled Gross-Pitaevskii equation and formula \eqref{eq:gamma-2-body}.  The calculations were performed for ${}^{87}$Rb atoms in energy levels $\ket{1} := \ket{F=2,\,m_{\rm F}=1}$ and $\ket{2} := \ket{F=1,\,m_{\rm F}=-1}$,  trapped in a spherical harmonic traps with trap frequency $\omega = 2\pi \times 100$Hz.
	The third rate,  $\gamma_{22}$, equals to zero. \label{fig:gammas}}
\end{figure}

The term $\chi \Sz^2$ in the Hamiltonian \eqref{eq:ham-oat-simple}, which is responsible for generation of entanglement, comes in fact from elastic collisions between atoms. It happens, however, with little probability, that the collision is not elastic -- in consequence of such inelastic collision both atoms are lost from the trap. The rate of two-body losses
	\begin{equation}
		\gamma_{\epsilon \epsilon'} = \frac{1}{2} K_{\epsilon\epsilon'}\int\, {\rm d}^3r |\psi_{\rm GPE}^{(\epsilon)}|^2|\psi_{\rm GPE}^{(\epsilon')}|^2
		\label{eq:gamma-2-body}
	\end{equation}
depends on the rate of the binary collisions, which stems from both  Bose-Einstein wave-functions. Atomic constants  $K_{\epsilon\epsilon'}$ are associated with probabilities that the collision will be inelastic.
In the limit of large number of atoms the condensates enter the so called Thomas-Fermi regime, in which one finds excellent analytical approximations for the solutions of the Gross-Pitaevskii equations.
Using this Thomas-Fermi approximations one may evaluate Eq. \eqref{eq:gamma-2-body} to find analytical approximation for two-body loss rates. In case of atoms in the spherically symmetric harmonic trapping potential, assuming equal coupling strengths $g_{11} = g_{22} = g_{12} =: g $ one gets
	\begin{equation}
	\gamma_{\epsilon\epsilon'}^{(2)} = \frac{(15)^{2/5}}{28\pi} \frac{K_{\epsilon\epsilon'}}{l_{\rm osc}^3} \bb{\frac{l_{\rm osc}}{a}}^{3/5} N^{-3/5},
	\label{eq:gamma-TF}
	\end{equation}
where $a = \frac{g \,m }{4\,\hbar^2\, \pi}$ is called scattering length, $l_{\rm osc} =\sqrt{\frac{\hbar}{m\omega}}$ is the oscillatory length and $N$ is the initial number of atoms. Due to the factor $N^{-3/5}$ the rate of two-body losses is getting lower when the number of atoms increases.  This is caused by the van der Waals repulsion between condensed atoms, which results in the spatial broadening of the condensate wave function. Finally, the integrals $\int\, {\rm d}^3r |\psi_{\rm GPE}^{(\epsilon)}|^2|\psi_{\rm GPE}^{(\epsilon')}|^2$  decays, so as the two-body rates as given in Eq. \eqref{eq:gamma-2-body}. This has far going consequences for scaling of the optimal Quantum Fisher Information $F_{\rm Q}$.
One has to remember that the total number of lost atoms per unit of time increases with the total number of atoms, but with slower rate compared to the case with constant $\gamma$.  In Fig. \ref{fig:gammas} the approximation \eqref{eq:gamma-TF} is compared to the exact result, obtained directly from  Eq. \eqref{eq:gamma-2-body}, but using the exact numerical solutions of equations \eqref{eq:GPE-coupled} for the orbitals $\psi_{\rm GPE}^{(\epsilon)}$.

\begin{figure}
\begin{center}
\includegraphics[width=\columnwidth]{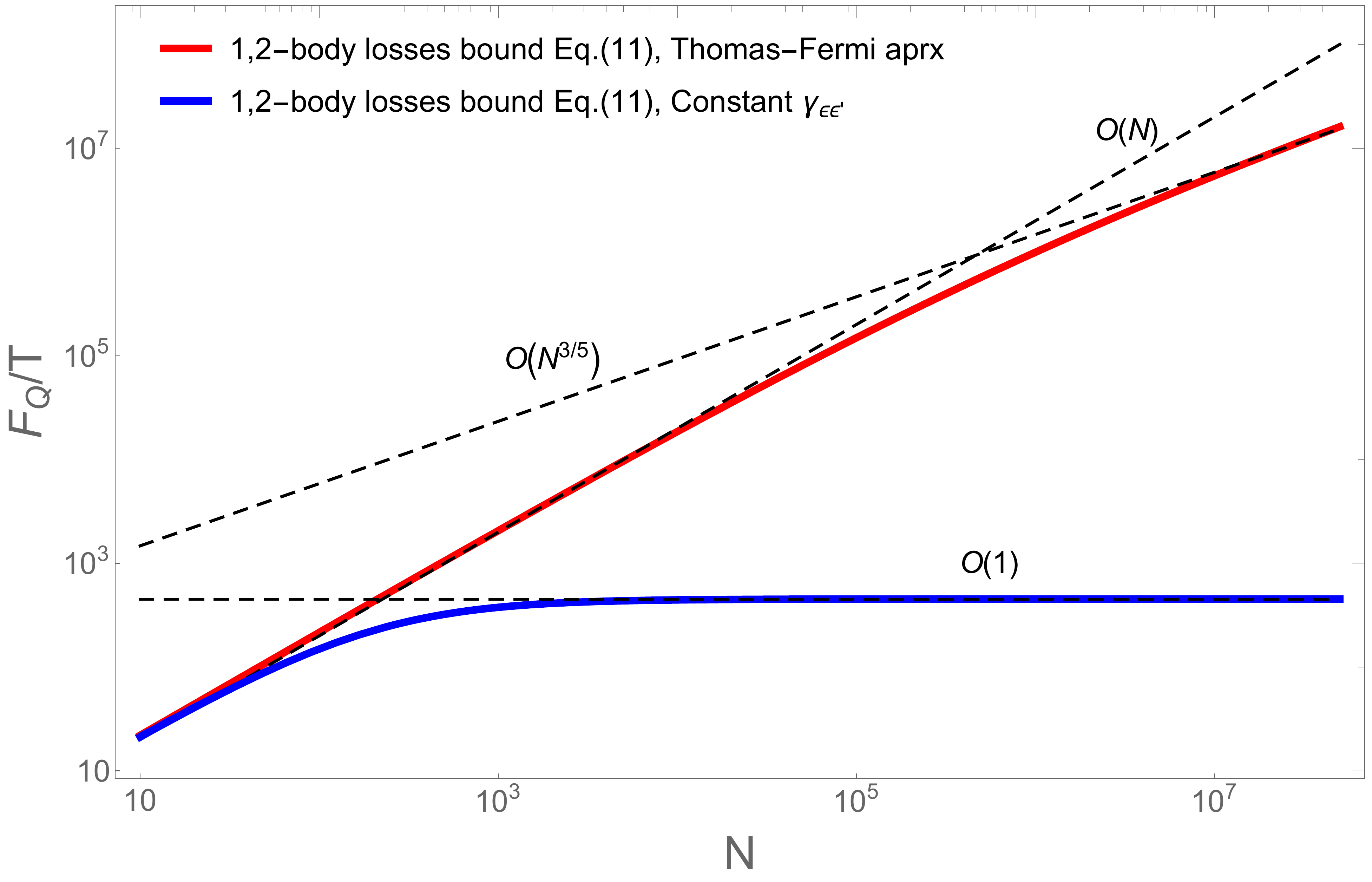}
\caption{Bounds on QFI in different models of scaling of two-body loss parameters with the number of particles $N$. The red curve is the bound Eq.~\eqref{eq:rpnbound} calculated using the Thomas-Fermi approximation Eq.~\eqref{eq:gamma-TF} and parameter values from the experiment of \cite{Ockelon2013}. The blue curve represents the same bound but using $\gamma_{\epsilon,\epsilon'}\sim \text{const.}$ in the number of particles.
Depending on how $\gamma_{\epsilon,\epsilon'}(N)$ behaves, QFI asymptotically is constant, the case discussed in Sec.~\ref{sec:asymptotic}, or scales as $O(N^{3/5})$ in the Thomas-Fermi approximation. The linear scaling $O(N)$ corresponds to the bound coming form single-body losses. \label{fig:qfin}}
\end{center}
\end{figure}

Majority of experiments is performed in the harmonic trap and within Thomas-Fermi regime. Still, there are experiments \cite{Hadzibabic2013, zwierleinBox} in which the potential trapping atoms is practically equal to a box. In this case the condensate wave-functions  are practically constants $\psi_{\rm GPE}=1/\sqrt{V}$, where $V$ is the volume of the trap. Then Eq. \eqref{eq:gamma-2-body} leads to rate of two-body losses independent on the number of particles.
The influence of the two cases---the Thomas-Fermi approximation and the box trap---on bounds on QFI is depicted in Fig.~\ref{fig:qfin}. This figure also depicts the main result of this paper concerning scaling of the fundamental bounds with the number of particles. Constant decoherence parameters result in constant precision bound, as we have already showed in Eqs.~\eqref{eq:fullaabound}, \eqref{eq:fqloss2full}. If on the other hand $\gamma$ scales with $N$ the fundamental bound changes significantly: to $O(N^{3/5})$, which shows that manipulation of density of the atomic cloud  offers trade-offs visible also in fundamental bounds.

Another experimental complication is the fluctuation of the total number of atoms in the initial state. Usually $N$ is a stochastic variable, with distribution close to the Poissonian one. Due to the term $\tilde{\chi} \hat{N} \,\Sz$, the fluctuations of total $N$ leads to a phase noise. This effect is known under the name "collisional shift".
	
As it is impossible to fully avoid particle losses, hence the entangled state has to be produced and used on faster time-scale than the ones connected with particle losses. This demands large value of the non-linearity parameter $\chi$.
From the equation for $\chi$ \eqref{eq:chis} and energy \eqref{eq:E-GPE} one finds the approximation
	\begin{align}
	\chi \approx &\frac{1}{2}g_{11}\int {\rm d}^3r |\psi_{\rm GPE}^{(1)}|^4 + \frac{1}{2}g_{22}\int {\rm d}^3r |\psi_{\rm GPE}^{(2)}|^4 \nonumber \\
	&- g_{12}\int {\rm d}^3r |\psi_{\rm GPE}^{(1)}|^2|\psi_{\rm GPE}^{(2)}|^2.
	\end{align}
In the case of the widely used ${}^{87}$Rb atoms, the interaction strength parameters $g_{\epsilon\epsilon'} $ are close to each other. In consequence, if atoms in $\ket{1}$ and $\ket{2}$ share a common spatial mode, i.e. $\psi_{\rm GPE}^{(1)} \approx \psi_{\rm GPE}^{(2)}$, then the nonlinearity practically vanishes $\chi \approx 0$. One of the way to increase $\chi$ is to separate both clouds using the state dependent trapping potentials \cite{riedel2010} and varying their centers $r_{\sigma}^{(\epsilon)}$ , the other -- to change one of the interaction strengths  using Feshbach resonances \cite{gross2010nonlinear}.
			
\begin{figure}
		\includegraphics[width=\columnwidth]{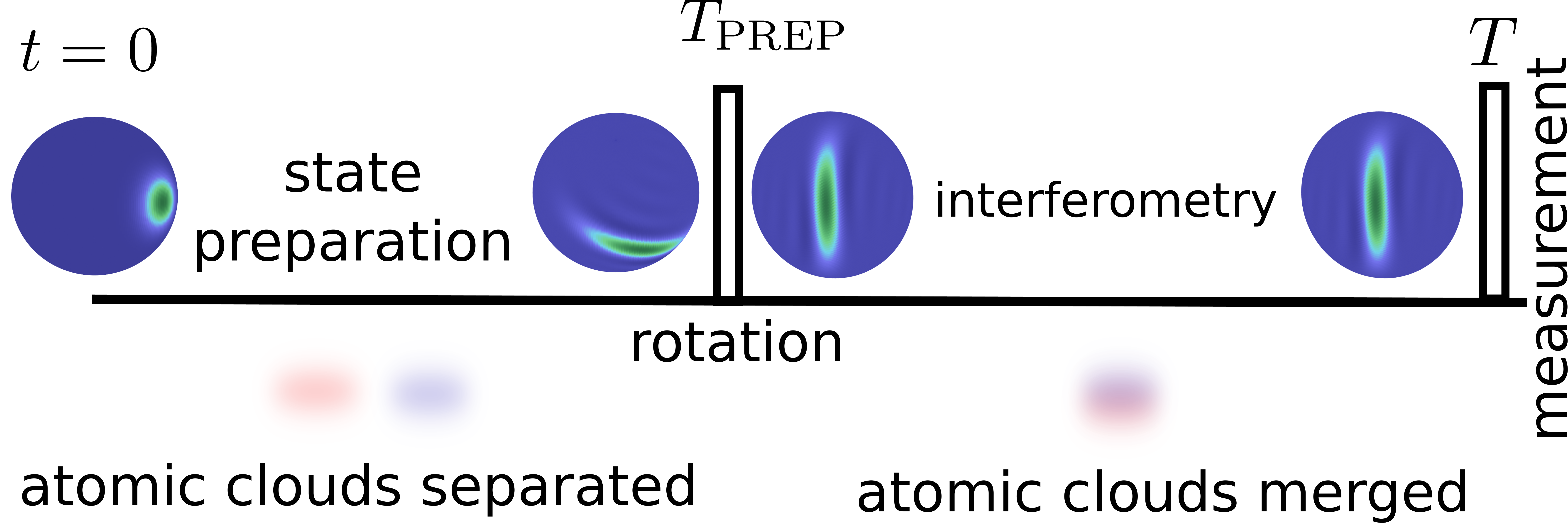}			
		\caption{Experimental scheme discussed in this paper. The initial state is a spin coherent state, as represented on the Bloch sphere with $\t{SU}(2)$ Wigner function. The entangling dynamics is initiated by taking the two atomic clouds apart to maximize non-linearity. In this preparation stage the useful state, as a spin squeezed state can be obtained.
		Due to asymmetry in the loss rates, the state leaves the equator as shown schematically in the pseudo Bloch sphere.
		To use the state in linear interferometry, first the two clouds have to be merged into a common spatial mode, and then the quantum state encoded in the internal degrees of freedom is appropriately rotated. Afterwards the state should accumulate a phase.
			\label{fig:scheme}}
	\end{figure}

In both cases, once the target entangled state is prepared at $T_{\rm PREP}$ during the nonlinear dynamics, the further entangling evolution has to be stopped. In the case of two separated condensates it is sufficient to bring both clouds together, ideally to a common spatial mode. The common spatial mode is also necessary to perform any $\t{SU}(2)$ rotations, usually required to adjust the state appropriately to the quantum information task. In particular,  Ramsey interferometry requires appropriate orientation of the state with respect to the phase-imprinting Hamiltonian. If phase will be accumulated in the dynamics generated by the Hamiltonian $\hat{H} \propto \Sz$, as discussed here, and the entangled state is squeezed, then the state on the Bloch sphere has to be elongated along the meridian, see Fig. \ref{fig:scheme}.
		
We were simulating numerically  the scheme presented in Fig. \ref{fig:scheme}.
For each chosen Ramsey time $T$, we were optimizing the following Fisher Information with respect to the preparation time $T_{\rm PREP}$
	\begin{equation}
	F_{\rm s} (T) =  \max_{T_{\rm PREP} \in [0, \infty]}\,\max_{0 < t<T} \bb{ \frac{T}{t}\,F_Q(t) }.
	\label{eq:FQ-short-bound}
	\end{equation}
The state of the system at time $t$ was computed from the master equations using quantum trajectory method \cite{dalibard1992}. The parameters of the master equations were computed numerically, by solving the coupled Gross-Pitaevskii equations \eqref{eq:GPE-coupled}, then evaluating Eqns. \eqref{eq:E-GPE}, \eqref{eq:chis} to find non-linearities $\chi$ and $\tilde{\chi}$ and \eqref{eq:gamma-2-body} to find the rate of two-body losses. Three body losses were also included, but they do not played any important role. The parameters were computed separately for the nonlinear evolution, during which the two condensates are taken apart, and separately for the interferometry , during which the two clouds should stay together (to avoid further nonlinear dynamics).
The initial number of atoms was stochastic with the Poissonian distribution to refer to the experimental situation.
As long as no special tricks are used, like the compensation method \cite{pawlowski2013, catparis}, then the optimal preparation times are short, resulting in the weakly squeezed states. In this case there is no substantial difference between the full Quantum Fisher Information and its estimate from the  error propagation formula based on the measurements of the spin components.


\begin{figure}
\begin{center}
\includegraphics[width=\columnwidth]{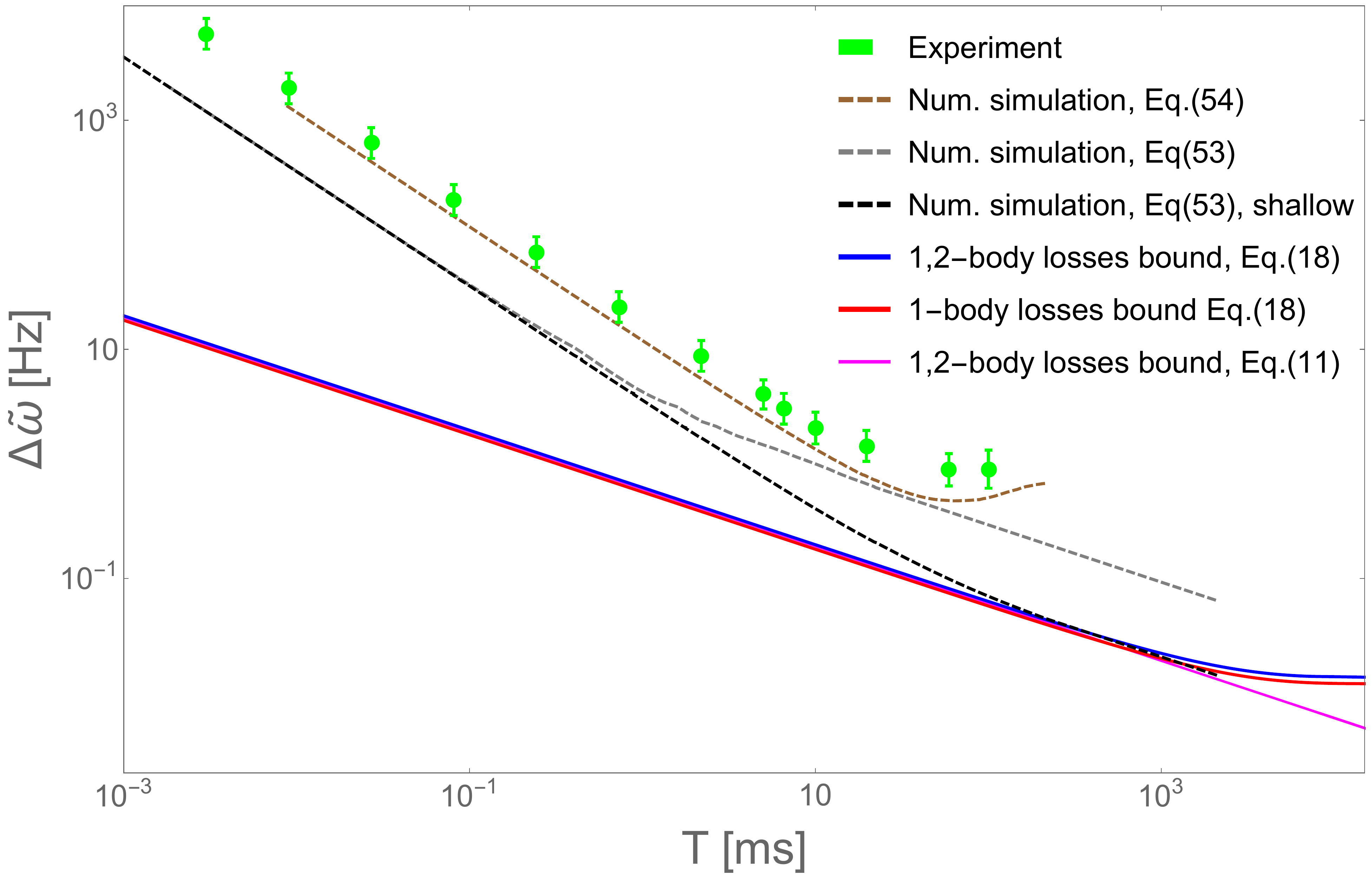}
\caption{Precision of estimation of $\omega$ in the case of both one and two body losses. We present theoretical bounds calculated according to Eq.\eqref{eq:rpntbound} for only one-body losses and for both types of decoherence present. Above the fundamental bounds we see the results of numerical simulations. Dashed curves show numerical simulation that includes time-optimization of Eq.~\eqref{eq:FQ-short-bound}, "shallow" denotes the simulation with a modulated trap that decreases the impact of two-body losses. The green points are the result of the experiment \cite{Ockelon2013}. In our numerical simulations we assumed the trap frequencies as in the experiment \cite{Ockelon2013}, equal to $(\nu_x, \nu_y, \nu_z) = (540, 540,115)$Hz with in average $N=1400$ atoms initially. The loss rates were computed from Gross-Pitaevskii equation applied to two clouds of $N_1=N_2=700$ atoms, specified in the caption of Fig. \ref{fig:gammas}. \label{fig:omegat}}
\end{center}
\end{figure}
In Fig.~\ref{fig:omegat} we present a comparison of precision of estimation coming from fundamental bounds, numerical simulations, and an actual experiment. The two curves representing fundamental bounds show that in the case of the experimental loss rates of  \cite{Ockelon2013} the influence of two-body losses on the best achievable precision is marginal. This small discrepancy between fundamental bounds is mainly caused by the small number of particles. Note that in Fig.~\ref{fig:qfin} for $N=1400$ the bound in the Thomas-Fermi approximation scales as QFI for single-body losses.
There is a subtle difference, between the experiment and theory: the experimental frequency estimation at time $T$, was not based on stopping the experiment at an intermediate time $t$ and then repeating the experiment $T/t$ times, as it is assumed in Eq.~\eqref{eq:FQ-short-bound}. The experimentally found $\Delta \omega$ should be rather benchmarked with $1/\sqrt{F_Q(T)}$, where $F_Q(T)$ comes directly from the definition \eqref{eq:qfi}. As in the following expression:
	\begin{equation}
	F_{\rm s} (T) =  \max_{T_{\rm PREP} \in [0, \infty]}\,  F_Q(t) .
	\label{eq:FQ-long-bound}
	\end{equation}
The deviations due to the subtleties in the definition of the uncertainty bound are definitely not sufficient to explain the huge difference between the experimental results and the ultimate bound.

We investigate the origins of the theory-to-experiment difference using the numerical simulations described above.
The brown curve in Fig.~\ref{fig:omegat} shows the numerical results for the trap geometry, the separation between clouds, and the fixed preparation time mimicking the experimental situation \cite{Ockelon2013}. On the other hand, the dynamical spatial phenomena  and the phase noise coming from the thermal effects are omitted in the simulation. Especially the former factor may be important---in the preparation stage of the very experiment \cite{Ockelon2013} the clouds' centers of masses were not stationary, but oscillating with amplitude around  $140$nm.
Although the simulation is not comprehensive, still it stays relatively close to the experiment.
Then we repeated the simulation but fully separating both clouds (which is still reasonable, compare with \cite{kurkjian2017}), optimizing over the preparation time and using the definition \eqref{eq:FQ-short-bound}. Due to larger separation the non-linearity parameter $\chi$ increases, whereas losses slightly decrease. By this, the initial states for Ramsey interferometry are less deteriorated by decoherence within the preparation stage.
Precision increases by order of magnitude. Especially important is the optimization over the preparation time---by adjusting  $T_{\rm PREP}$ separately to each Ramsey time $T$ we have found that experimentally strongly squeezed states are in fact a bad choice for long interferometry durations.
Finally we considered the next scenario,  with the atomic clouds placed for the Ramsay interferometry into a very shallow trap.
In this case, there are no two-body losses, neither the residual non-linearities effect, i.e. $\tilde{\chi}=0$, within interferometry.
This last simulation, shows how much one can gain by avoiding the two-body losses and the collisional shift in the real experiment.


\section{Conclusions\label{sec:conclusions}}
We have shown how to use the recently developed tools of theoretical quantum metrology to analyze in great detail interferometric experiments performed with cold atoms. We provide the fundamental and more realistic bounds on precision and complete the ongoing discussion on achievable scaling of QFI with the number of particles.
Future work might involve adopting presented methods in multi-parameter metrology. On the other hand, as the field of experimental non-linear metrolgy grows rapidly, it will be necessary to perform similar analysis for setups involving the quadratic Hamiltonian.
We analyzed theoretical bounds in the context of interferometry based on Bose-Einstein condensates. We benchmarked our theoretical results with experimental data \cite{Ockelon2013} and numerical simulations.
First, we have shown that our simulations are close to the experimental results presented in \cite{Ockelon2013}.
Then, we focused on ways to improve experimental precision by tuning "simple" parameters, like separation between atomic clouds in the preparation stage, duration of the entangling dynamics and frequency of the final trap. One could in principle increase the precision by an order of magnitude. For longer Ramsey times, of the order of $100$ms, it is actually beneficial to stop the "state preparation" stage early, much before reaching the maximally squeezed states. Numerical analysis should be extended by a better model of the spatial dynamics and the non-adiabatic effects -- the factors influencing strongly the quantum engineering using the Bose-Einstein condensates \cite{yun2009}.
As the benchmark between simulations for the optimized parameter and fundamental bounds shows a big room for improvement, one should perform full optimization of the tunable parameters to either direct the experiments on the fastest track or to find the most important limiting factors.

\section{Acknowledgments}
We thank Philipp Treutlein for giving us access to cold atom magnetometry experimental data.
JC acknowledges support by the Netherlands Organisation for Scientific Research (NWO) VIDI grant (Project No. 639.022.519).
RDD acknowledges support from the  National Science Center (Poland) grant No. 2016/22/E/ST2/00559.
KP acknowledges support from the National Science Center (Poland) grant No. 2014/13/D/ST2/01883.

\bibliography{atoms}

\newpage
\end{document}